\documentclass[11pt]{JHEP3}
\usepackage{epsfig}
\usepackage{latexsym}
\usepackage{graphicx}
\usepackage{amsmath,amssymb}

\def\bea{\begin{eqnarray}}
\def\eea{\end{eqnarray}}
\def\be{\begin{equation}}
\def\ee{\end{equation}}

\def\Z{{\bf Z}}

\def\b{\raisebox{14pt}{}\raisebox{-7pt}{}$\!\!$}

\newcommand{\into}{\rightarrow}

\newcommand{\bnb}{^{\!\!\!\!\!\!\hbox{\tiny{(\;\,\,)}}}}
\newcommand{\bnbs}{^{\!\!\!\!\!\hbox{\tiny{(\;\,\,)}}}}
\newcommand{\ds}{{\sffamily DarkSUSY}}

\title{A Dark Matter Candidate from an Extra (Non-Universal) Dimension}
\author{Marco Regis, Marco Serone and Piero Ullio\\
International School for Advanced Studies (SISSA/ISAS)
and INFN, Trieste.\\
E-Mail: \email{regis,serone,ullio@sissa.it}}

\abstract{
We show that a recently constructed five--dimensional (5D) model with gauge--Higgs unification and explicit Lorentz
symmetry breaking in the bulk, provides a natural dark matter candidate.
This is the lightest Kaluza--Klein particle odd under a certain discrete $\Z_2$ symmetry,
which has been introduced to improve the naturalness of the model, and
resembles KK--parity but is less constraining.

The dark matter candidate is the first KK mode of a 5D gauge field and
electroweak bounds force its mass above the TeV scale.
Its pair annihilation rate is too small to guarantee the correct relic abundance;
however coannihilations with colored particles greatly
enhance the effective annihilation rate, leading to realistic relic densities.
}

\preprint{SISSA-83/2006/EP}


\begin{document}

\section{Introduction}

The latest years have been marked by the tremendous progresses in observational cosmology.
As cornerstones, the detailed maps of cosmic microwave background~\cite{Spergel:2006hy}
and of the three-dimensional distribution of galaxies in the Universe~\cite{Tegmark:2006az}
have allowed very significant improvements in the discrimination among cosmological models
and in the determination of cosmological parameters. In particular, the case for non-baryonic
dark matter (DM) as building block of all structures in the Universe has become stronger and
stronger: its contribution to the present mean energy density is found to be
$\Omega_{\rm DM} h^2 = 0.105 \pm 0.004$~\cite{Tegmark:2006az} (as usual, in this formula
the mean DM density is normalized to the critical density
$\rho_c = 1.879 \times 10^{-29} h^2 {\rm g}/{\rm cm}^3$, with $h = 0.730 \pm 0.019$ being the
Hubble constant in units of $100\,{\rm km}\,{\rm s}^{-1}\,{\rm Mpc}^{-1}$).
The nature of the DM is still unknown. Among viable scenarios, thermal generation
seems the most natural DM production mechanism, and weakly interacting massive particles (WIMPs)
are among the leading DM candidates: since they are massive, their decoupling from thermal
equilibrium occurs in the non-relativistic regime; the weak interaction rate with lighter
standard model (SM) particles ensures that their thermal relic density is naturally of the order of
$\Omega_{\rm DM}$~\cite{Lee:1977ua,Gunn:1978gr} (for reviews on dark matter candidates,
see e.g.~\cite{review1,review2}).

Essentially all theories in extension to the SM predict the existence of new massive
particles;  some of this extra states can indeed be ``dark", i.e. be color and electromagnetic
neutral, with the weak force (and gravity) as relevant coupling to ordinary matter.
A better understanding of the mechanism of Electroweak Symmetry Breaking (EWSB)
is one of the strongest motivations to consider models beyond the SM (BSM);
it is indeed tempting to search for a framework embedding, at the same time, naturalness
for EWSB and WIMPs as DM constituents of the Universe. This is plausible whenever
there is a mechanism preventing the WIMP to decay (or forcing its lifetime to be much
longer than the present age of the Universe). The condition of stability is usually
fulfilled by introducing a new unbroken discrete $\Z_2$ symmetry: all SM particles are
assumed to be neutral under this symmetry, while the WIMP DM candidate is the lightest
non--neutral state. Relevant examples of BSM theories which aim to resolve or alleviate
the SM instability of the EWSB and provide DM candidates, include supersymmetric and little
Higgs theories; in these two cases, the $\Z_2$ symmetry is  identified, respectively, with
the R-parity~\cite{Farrar:1978xj} and the
T-parity~\cite{Cheng:2003ju}.

Higher dimensional theories may fit as well into this picture: the lighest Kaluza--Klein
particle (LKP) is potentially a good DM candidate in the class of extra dimension scenarios
in which a discrete symmetry makes the LKP stable.
The simplest models are 5D theories with Universal Extra Dimensions (UED)~\cite{Appelquist:2000nn},
namely theories where all the SM particles are promoted to bulk fields propagating in
higher dimensions, where such a symmetry is the so-called KK-parity,
an unbroken $\Z_2$ subgroup of the translation group in the extra dimensions~\cite{Servant:2002aq}  (see
e.g.~\cite{Allanach:2006fy} for other frameworks arising from extra dimensions).
Despite the simplicity of these models, however, UED theories do not shed
any light on the EWSB mechanism of the SM, whose quantum instability gets actually worse
because of the higher (cubic) dependence of the Higgs mass on the UV cut--off of the theory.

One of the main results in this work is to show that stable DM candidates can be embedded
also in non-universal higher dimensional theories aiming at the stabilization of the electroweak
scale. For such purpose, we will focus in a recently proposed 5D theory in which the Higgs field
is the internal component of a gauge field, and Lorentz symmetry is broken in the
bulk~\cite{Panico:2006em}  (see e.g. \cite{Serone:2005ds} for a brief pedagogical review
of such kind of models and for further references). Within this framework, a $\Z_2$ symmetry (called mirror
symmetry) has been invoked to improve the naturalness of the model~\cite{Panico:2006em};
as a by-product, this symmetry guarantees  the stability of the lightest $\Z_2$ odd particle.
$\Z_2$ symmetries of this kind
are less restrictive than KK--parity. Their implementation is particularly intuitive if one considers
5D theories on an interval $S^1/\Z_2$.  The mirror symmetry acts on a given field and its copy
under the symmetry, giving rise to periodic and anti-periodic states along the covering circle
$S^1$, respectively even and odd under the mirror symmetry. The LKP is then identified
with the first KK mode of the lightest 5D antiperiodic field in the model, similarly to the LKP in
UED models, but with the important difference that mirror symmetry is not a remnant of a
space-time symmetry and hence does not necessarily act on all fields in the model. In particular,
the mirror symmetry we propose here can be implemented in  flat as well as warped spaces, and
does not put any constraint on the relation between the boundary Lagrangians at the two
fixed-points, aside the obvious one of being $\Z_2$ even.

We present here a detailed calculation of the thermal relic density of the LKP in the model
of~\cite{Panico:2006em}. Since Lorentz symmetry in the extra dimension is explicitly broken,
there is a certain degree of uncertainty in the model mass spectrum.
We focus on the region in the parameter space where the LKP is the first KK
mode of an antiperiodic gauge field, roughly aligned along the $U(1)_Y$ direction in field
space. Electroweak bounds require this field to be heavier than about 2~TeV, in
a range which is significantly more massive than the analogous state in the UED
scenario~\cite{Servant:2002aq}, as well as most WIMP DM candidates. Since the mass is so
heavy, the pair annihilation rate for our WIMP candidate is small and would tend to lead to the
departure from thermal equilibrium at too early times, overproducing DM by one order of
magnitude or more. On the other hand, the LKP appears within a set of other extra
antiperiodic fields, most often with the next-to-lightest Kaluza--Klein particle (NLKP) being
a strongly interacting particle. For reasonable values of parameters in the model, the mass
splitting between NLKP and LKP turns out to be small, and the NLKP becomes the particle
triggering the freeze-out and possibly lowering the LKP relic density within the observed value
(these are known as coannihilation effects~\cite{Griest:1990kh}). In particular, the nature of
the EWSB in the model implies that typically the lightest $\Z_2$--odd fermion is the
$b_-$, arising from the KK tower associated to the bottom quark. A strongly-interacting NLKP
gauge boson can be found, instead, in case the mirror symmetry acts on the color $SU(3)_s$.
For simplicity, we then discuss two classes of viable scenarios:
\begin{enumerate}
\item{The LKP coannihilates with the $b_-$, and gluons are periodic on $S^1$.}
\item{Gluons are both periodic and antiperiodic on $S^1$ and the LKP coannihilates also with the
first KK mode of the antiperiodic gluon.}
\end{enumerate}
Note that in the first scenario there is  a further increase in the effective thermally averaged
annihilation cross section due to a KK-gluon s-channel resonance in $b_-$ pair annihilations.
Values of the relic density in agreement with observations are obtained in both scenarios,
with a moderate degree of fine-tuning (of order few percent), comparable or even lower than
what one obtains
in other cases in the literature
when the relic density of the WIMP DM candidate is driven by coannihilation effects.

The structure of the paper is as follows. In Section 2, we introduce the mirror symmetry and briefly
review the essential ingredients of the model \cite{Panico:2006em}, focusing in particular to the mass spectrum of the lightest states. In Section 3 we compute the relic density for the two scenarios mentioned above and we add some remarks
about the fine--tuning needed to get the correct relic density.
Section 4 concludes. Various details regarding the Feynman rules in our model, a one--loop mass splitting computation, the list of all processes relevant for the relic density calculation, and
the running of the strong coupling constant $\alpha_s$
are contained in the appendices.

\section{Mirror Symmetry and a DM Candidate}

An interesting property of models based on Universal Extra Dimensions (UED) \cite{Appelquist:2000nn}
is the possible presence of a $\Z_2$ symmetry, remnant of the broken translations along the extra dimension,
called KK--parity. This symmetry is crucial to make stable the lightest KK particle (LKP) and to identify
it as a suitable DM candidate \cite{Servant:2002aq}.
KK--parity inverts the segment around its middle point. In terms of a coordinate $0\leq y \leq \pi R$,
it implies the invariance of the Lagrangian under the transformation $y\rightarrow \pi R - y$.
Such invariance implies, in particular, the equality of any possible localized Lagrangian terms at $y=0$
and at $y=\pi R$: ${\cal L}_0={\cal L}_\pi$. Most extra dimensional models which aim
to stabilize in one way or another the electroweak scale, however, requires ${\cal L}_0\neq {\cal L}_\pi$
and do not respect KK--parity. In particular, models based on 5D warped spaces \cite{Randall:1999ee} manifestly
violate this symmetry. It is then desirable to impose some other less constraining symmetry protecting
some KK modes from decaying.

The $\Z_2$ symmetry we will consider below has been introduced in \cite{Panico:2006em} and allows for arbitrary localized terms in the Lagrangian. As it will be clear below, it works for both flat and warped spaces.
Consider a simple toy model of two interacting 5D real scalar fields $\phi_1$ and $\phi_2$
and impose on the system a $\Z_2$ symmetry which interchange them: $\phi_{1,2}\leftrightarrow \phi_{2,1}$.
Being the Lagrangian invariant under this symmetry, we can impose boundary conditions of the following form
for $\phi_1$ and $\phi_2$ (in the $S^1/\Z_2$ orbifold notation):
\be
\phi_1(y+ 2\pi R) =  \phi_2(y) \,,
\;\;\;\;\;\;\;\phi_{1}(-y) =  \eta \phi_{2}(y) \,,
\label{bound-cond}
\ee
where $\eta = \pm$. It is convenient to define linear combinations $\phi_{\pm}=(\phi_1\pm \phi_2)/\sqrt{2}$
which are respectively periodic and antiperiodic on the covering circle $S^1$ and with definite
orbifold parities: $\phi_\pm(-y) = \pm \eta \phi_\pm(y)$. Equivalently, one can consider in Eq.~(\ref{bound-cond})
the standard parity projection $\phi_1(-y)=\eta \phi_1(y)$ instead of $\phi_{1}(-y) =  \eta \phi_{2}(y)$,
resulting in a change of parity for $\phi_-$.
Under the $\Z_2$ symmetry, $\phi_{\pm}\rightarrow \pm \phi_{\pm}$, so we can assign a multiplicative charge $+1$ to $\phi_+$ and $-1$ to $\phi_-$.
The localized Lagrangian terms ${\cal L}_0$ and ${\cal L}_\pi$, which can include boundary fields as well,
can be arbitrary and in general different
from each other, provided they respect the above $\Z_2$ symmetry. We denote such $\Z_2$ symmetry
as ``mirror symmetry'' in the following.
It can also be implemented on gauge fields. For abelian gauge groups, it works as before and one is
left with two gauge fields, one periodic and one anti--periodic. For non-abelian gauge groups, mirror symmetry
can be easily implemented only when the orbifold twist (or the boundary conditions on the segment) are
trivially embedded in the gauge group. In such a case, starting from two identical gauge groups
$G_1\times G_2$,
the boundary conditions (\ref{bound-cond}) leave unbroken in 4D only the diagonal subgroup
$G_+$.\footnote{Notice that the antiperiodic gauge fields $A_-$ are {\it not} connections of the gauge
group $G_-$. The latter is not a group, but the symmetric quotient $(G_1\times G_2)/G_+$.}
Mirror symmetry changes the sign of all half--integer KK modes, associated to the antiperiodic field
$\phi_-$, leaving invariant the integer KK modes of $\phi_+$. As such, the first half-integer $n=1/2$
KK mode of $\phi_-$ cannot decay and is stable. Mirror symmetry acts on these fields as KK--parity, provided
one rescales $R\rightarrow R/2$, but with the important difference, as already
pointed out, of allowing more freedom in the 5D theory and on the localized 4D Lagrangian terms.

It is straightforward to generalize the action of mirror symmetry for more extra dimensions.
For instance, for complex scalars $\phi_1$ and $\phi_2$ compactified on a $T^2/Z_2$ orbifold
one can have
\be
\phi_1(z+1) =  \phi_2(z) \,,
\;\;\;\;\;\;\;\phi_{1}(z+\tau) =  \phi_{2}(z) \,, \;\;\;\;\;\;\;\; \phi_1(-z)=\eta \phi_2(z)\,,
\label{bound-cond-6}
\ee
with $z$ properly normalized dimensionless coordinates on $T^2$ and $\tau$ its modular parameter.
As in the 5D case, the lowest KK mode of $\phi_-$ is absolutely stable.

{} From a model-building point of view, it is of course desirable not to impose
mirror symmetry {\it ad hoc} for the only purpose of getting a stable particle, possibly with the correct properties
of being a good DM candidate. This is not mandatory but makes the symmetry ``more natural''.
In Supersymmetry, for instance, R--symmetry is typically imposed not only
to have a stable supersymmetric particle but also to avoid baryon--violating operators that would
lead to a too fast proton decay.\footnote{See \cite{Agashe:2004ci} for a 5D warped model where a $\Z_3$ discrete symmetry
is imposed to both suppress proton decay and have a stable non-SM particle.}
In the following, in the same spirit, we will consider a model \cite{Panico:2006em}
where mirror symmetry has been introduced to reduce the fine-tuning needed to stabilize the electroweak scale.

\subsection{A Specific Model}

The model we consider is a model of gauge-Higgs unification on a flat 5D space of the form
$R^{1,3}\times S^1/\Z_2$. It is well known that in models of this sort is hard to get
sufficiently heavy masses for the Higgs field and the top quark, due to various symmetry
constraints, including 5D Lorentz symmetry. The latter symmetry, in particular,
links gauge and Yukawa couplings between each other and does not easily allow to get the correct
top Yukawa coupling. Due to the radiative origin of the Higgs potential, a large Yukawa
coupling will also tend to increase the Higgs mass.
It has been shown in \cite{Panico:2005dh,Panico:2006em} that
by explicitly breaking 5D Lorentz symmetry in the bulk (leaving the 4D Lorentz symmetry unbroken),
one can easily overcome the two above problems of too light Higgs and top fields, having now no constraint
linking gauge and Yukawa couplings.
In the following, we review very briefly the main features of the model --- referring the interested reader
to \cite{Scrucca:2003ra,Panico:2005dh,Panico:2006em} for further details ---
and then consider in some detail the mass spectrum of the lightest non-SM states.

The gauge group is taken to be of the form $G \times G_1 \times G_2$,
with a certain number of couples of bulk fermions $(\Psi_1,\widetilde \Psi_1)$
and $(\Psi_2,\widetilde \Psi_2)$, with identical quantum numbers under the group $G$
and opposite orbifold parities.
We require that the Lagrangian is invariant under the mirror symmetry
$1\leftrightarrow 2$.
The couples $(\Psi_1,\widetilde \Psi_1)$ are charged under $G_1$ and neutral under $G_2$ and, by mirror symmetry, the same number of couples $(\Psi_2,\widetilde \Psi_2)$ are charged under $G_2$ and neutral under $G_1$.
No bulk field is simultaneously charged under both $G_1$ and $G_2$.

We can make two different choices for $G$ and $G_{1,2}$, depending
on whether we double the color group or not.
We can either take $G=SU(3)_w\times SU(3)_s$ and $G_i = U(1)_i$ or
$G=SU(3)_w$ and $G_i = SU(3)_{i,s}\times U(1)_i$ ($i=1,2$).\footnote{The doubling of the $U(1)$ factor is
necessary and motivated by naturalness \cite{Panico:2006em}.} As we will see, both choices
can give rise to a DM candidate with the correct relic density.
For definiteness, we focus in the following on the case in which $G_i = SU(3)_{i,s}\times U(1)_i$;
the other case can be trivially derived in analogy.
In total, we introduce (for each SM generation) one pair of
couples $(\Psi_{1,2}^{u},\widetilde \Psi_{1,2}^{u})$ in the
anti-fundamental representation of $SU(3)_w$ and one pair of couples
$(\Psi_{1,2}^{d},\widetilde \Psi_{1,2}^{d})$ in the symmetric representation
of $SU(3)_w$. Both pairs have $U(1)_{1,2}$ charge +1/3 and are in the fundamental representation
of $SU(3)_{1,2,s}$. The boundary conditions of these fermions and gauge fields follow from
Eqs.~(\ref{bound-cond}) and the twist matrix introduced in \cite{Scrucca:2003ra}.
The unbroken gauge group at $y=0$ is $SU(2) \times U(1) \times G_+$,
whereas at $y=\pi R$ we have $SU(2) \times U(1) \times G_1\times G_2$.
We also introduce massless chiral fermions with the SM quantum numbers and $\Z_2$ charge +1,
localized at $y=0$.
Mirror symmetry and the boundary conditions (\ref{bound-cond})
imply that the localized fields can (minimally) couple only to $A_+$ and mix with the
bulk fermions $\Psi_+$.

Before EWSB, the massless bosonic 4D fields are the gauge bosons
in the adjoint of $SU(2)\times U(1)\subset SU(3)_w$, $U(1)_+$, gluon gauge fields $g_+$
and a charged scalar doublet Higgs field, arising from the internal
components of the odd $SU(3)_w$ $5D$ gauge fields.
The $SU(3)_{+,s}$ and $SU(2)$ gauge groups are identified respectively
with the SM $SU(3)_s$ and $SU(2)_L$ ones, while the hypercharge $U(1)_Y$ is
the diagonal subgroup of $U(1)$ and $U(1)_+$.
The extra $U(1)_X$ gauge symmetry surviving the orbifold
projection is anomalous and its corresponding zero mode gauge boson gets a mass of the order of the cut-off scale
$\Lambda$ of the model ($\Lambda\simeq (3\div 4)/R$ \cite{Panico:2006em}).
The massless fermionic 4D fields, identified with the SM fermions, are the zero modes of a coupled
bulk--to--boundary fermion system.
Differently from the bosonic massless fields above, which all have a constant profile
along the fifth dimension, fermions have a profile which depends on the bulk--to--boundary
mixing terms. To a reasonable approximation, one can consider all SM fermions localized at $y=0$,
with the exception of the bottom quark, which shows a small wave-function tail away from $y=0$ and
the top quark, which is nearly totally delocalized.
All SM fields are even under mirror symmetry with the lightest $\Z_2$ odd
state in the model absolutely stable.
Since the bulk fermions $\Psi_{\pm}$ have 5D Dirac mass terms, in a (large) fraction of
the parameter space of the model, as we will see below,  the lightest $\Z_2$ odd state is the first KK mode of
the antiperiodic $U(1)_-$ gauge field, denoted by $A_-$.

\subsection{Mass Spectrum}

Electroweak constraints fix the compactification scale
in the multi-TeV regime. More precisely, it has been found in \cite{Panico:2006em} that
$1/R\geq 4.7$ TeV at 90$\%$ C.L. to pass all flavour and CP conserving bounds.
The lightest non-SM particles turn out to be in the 1 TeV range and thus
for all practical purposes we can neglect EWSB effects and consider the mass
spectrum in the unbroken phase.

Let us first consider $\Z_2$ even gauge bosons.
Aside from the massless SM fields considered before, we have a standard tower of KK states for all
gauge fields, with the exception of $X$, the gauge field of the anomalous $U(1)_X$ symmetry,
which becomes effectively a field with Dirichlet/Neumann boundary conditions at $y=0/\pi R$ and of $Y$,
the gauge field of the hypercharge $U(1)_Y$, which
can mix with $X$. We have then ($n\geq 1$),
\bea
m_{W_+}^{(2n)} & = & \frac{n}{R}\,, \label{MS-A+} \\
m_{g_+}^{(2n)} & = & \rho_s \frac{n}{R}\,, \label{MS-g+}
\eea
where $m_{W_+}^{(2n)}$ and $m_{g_+}^{(2n)}$ denote the masses of all $SU(3)_w\times U(1)^\prime$ and
$SU(3)_{+,s}$ gluon KK gauge fields except $X$ and $Y$. Since Lorentz invariance is broken in the bulk,
we have in general introduced the Lorentz--violating parameters $\rho$ and $\rho_s$,
which are the coefficients for the gauge kinetic terms of the form $F_{\mu 5}^2$ for $U(1)^\prime$ and
$SU(3)_s$ respectively (see \cite{Panico:2006em} for further details). In the following, we will mostly consider
the case in which $\rho\sim \rho_s \simeq 1$, the Lorentz--invariant value.
When $\rho\simeq 1$, the mixing between $Y$ and $X$ is negligible and their KK masses are given by
\bea
m_{Y_{+}}^{(2n)} & \simeq & \rho\frac{n}{R}\,, \label{MS-Ay+} \\
m_{X_{+}}^{(2n)} & \simeq & \rho \frac{(n-1/2)}{R}\,. \label{MS-Ax+}
\eea
The mass spectra of $\Z_2$ odd gauge bosons is easily derived, since no anomalies arise here. We have
\bea
m_{g_-}^{(2n-1)} & = & \rho_s \frac{(n-1/2)}{R}\,, \label{MS-g-}\\
m_{A_-}^{(2n-1)} & = & \rho \frac{(n-1/2)}{R}\,. \label{MS-A-}
\eea

The mass spectra for periodic $SU(2)_L$--triplet fermions and for all antiperiodic fermions is also easily computed,
since they cannot mix with boundary fermions. One has
\be
\begin{cases}
m_{i+}^{(2n)}= \sqrt{M_i^2 + k_i^2\left(\frac{n}{R}\right)^2} & n\geq 0\\
m_{i-}^{(2n-1)}= \sqrt{M_i^2 + k_i^2\left(\frac{(n-1/2)}{R}\right)^2}\,\,\,\,\,& n\geq 1\,\,,
\end{cases}
\ee
where $k_i$ are the Lorentz-violating factors entering in the covariant derivative of the fermions and $M_i$
are bulk mass terms (notation as in \cite{Panico:2006em}).

The mass spectra for $SU(2)_L$ doublet and singlet periodic fermions is more complicated and
given by the roots of transcendental equations
which do not admit simple analytic expressions. These equations depend on
the bulk--to--boundary mixing terms $\epsilon_{1,2}^i$, the parameters $k_i$
and the bulk mass terms $M_i$.
After EWSB, the SM fermion masses are function of these parameters, so that the subspace
of the parameter space spanned by $(\epsilon_i,k_i,M_i)$ is not totally arbitrary.
In addition, the electroweak constraints, perturbativity and an estimate of the size
of possible Flavour Changing Neutral Currents (FCNC) favours a given regime for such parameters.
For all quarks and leptons, except the top and bottom quarks, $\epsilon_{1,2}^i\simeq 0.1$, $k_i\simeq 1$. For the
bottom quark we have $\epsilon^b_{1,2}\simeq 0.2$, $k_b \simeq 1$ and for the top quark
$\epsilon^t_{1,2}\simeq 1.2$, $k_t \simeq 2.5$\footnote{This is the only
needed and relevant Lorentz violating coupling in the model.}. Having fixed $\epsilon_{1,2}^i$
and $k_i$, the bulk mass terms $M_i$ are derived by the known values of the SM fermion masses.

\begin{figure}
\centering
\includegraphics[width=10cm]{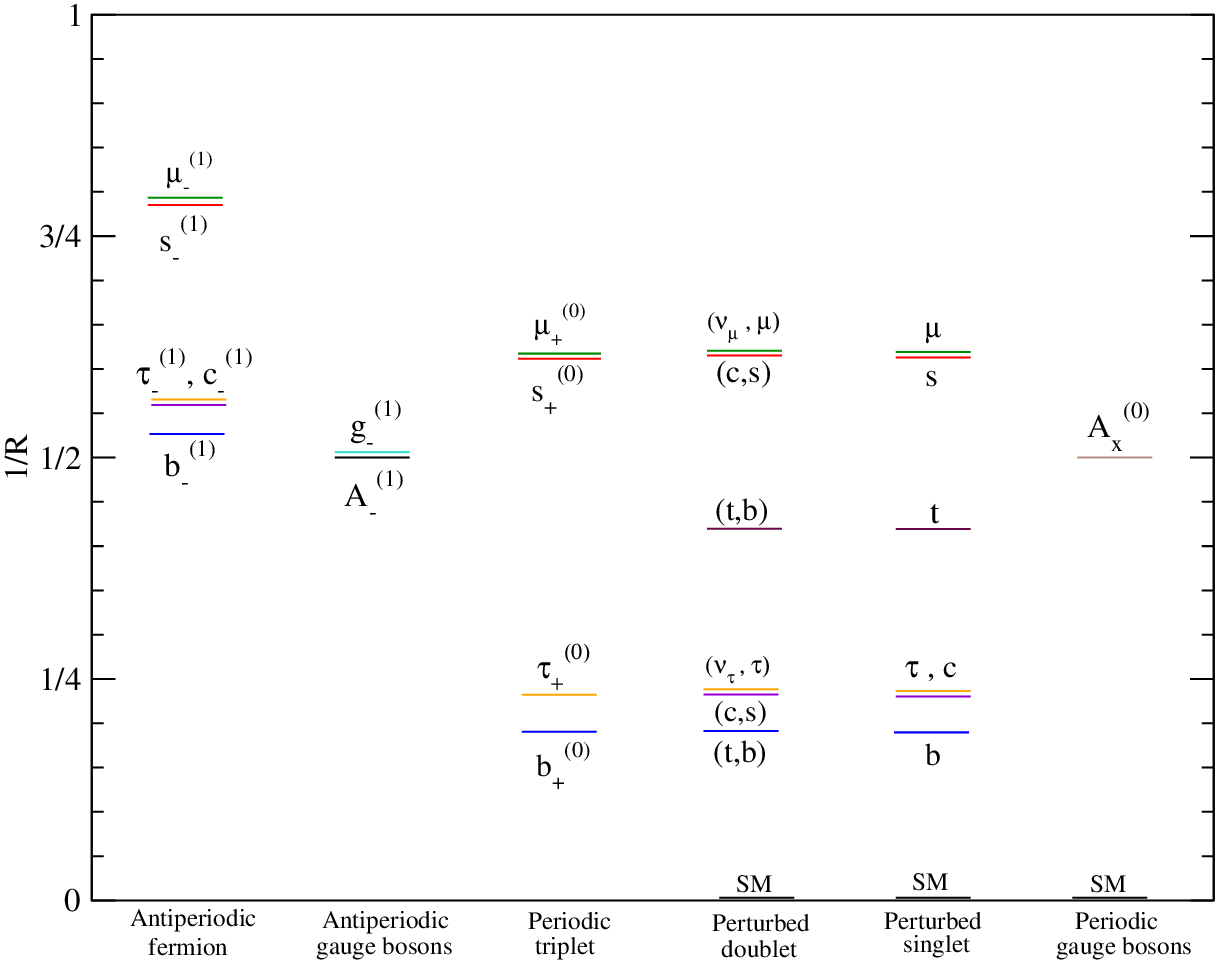}
\caption{Tree--level spectrum for all states with mass $< 1/R$. The DM candidate is $A_-^{(1)}$.}
\label{figmass}
\end{figure}

We summarize in Fig.~\ref{figmass} the masses of the lightest KK states for the typical values
of the parameters considered above. We report the tree--level mass spectra for both $\Z_2$ odd and even states for completeness, although the latter do not play an important role in the thermal relic density computation.
We denote by $b_-^{(1)}, c_-^{(1)}$, etc. the first $n=1$ KK mode of the corresponding antiperiodic fermions
$\Psi_-^b,\Psi_-^c$ and so on. Similarly, for the $n=0$ KK modes $b_+^{(0)}$, $c_+^{(0)}$, etc. of the $SU(2)_L$ triplet
fermions. The fields in the fourth and fifth column in Fig.\ref{figmass} (perturbed doublet and perturbed singlet)
are the first periodic massive resonances of the corresponding SM fields.
For $\rho_s\gtrsim \rho$, the lightest $\Z_2$ particle is the first
$n=1$ mode of $A_-$, denoted by $A_-^{(1)}$, which will be our DM candidate.\footnote{The DM candidate might
also be identified with an unstable, but sufficiently long--lived, particle.
In extra dimensions, a candidate of this sort might be the radion, whose relic abundance is typically
too large \cite{Kolb:2003mm}.
In our scenario, most likely the radion physics
will be entangled with the microscopic mechanism inducing the 5D Lorentz breaking,
which might also provide a stabilization mechanism for the radion.
The radion physics should then be revised. This analysis is beyond the aims
of our paper and may deserve further study.} As can be seen from Fig.~\ref{figmass},
it does not coincide with the lightest non-SM particle in the model,
the latter being given by two $\Z_2$--even fermions, $SU(2)_L$ triplets,
which are almost degenerate with an other $\Z_2$ even fermion, $SU(2)_L$ singlet.
They all come from the KK tower associated to the bottom quark and have a mass $\sim  1/(5R)$.

Having various free parameters governing the masses of the relevant KK modes, it is pointless
to compute the mass corrections induced by the EWSB and radiative corrections.
They can all be encoded in the effective parameters $\rho$, $\rho_s$ and $k_i$.\footnote{As we will see in the
following, the region in parameter space where $\rho_s \simeq \rho$ is the most interesting as far as DM
is concerned. Strictly speaking, then, we are considering tree-level values of $\rho_s$ and $\rho$
which differ by the correct amount to compensate the splitting induced by quantum corrections.}
There is however a case in which radiative corrections are relevant and need to be computed.
When the $n=1$ KK gluons $g_-^{(1)}$ (or KK fermions $b_-^{(1)}$) coannihilate with $A_-^{(1)}$, the s--channel diagram in which
a $g_+^{(2)}$ is created in the $g_-^{(1)}\!\!-\!g_-^{(1)}$ (or $b_-^{(1)}\!\!-\!b_-^{(1)}$) annihilation might be in resonance
and amplify the annihilation in question, decreasing the relic density.
Although the absolute radiative correction to the mass of $g_-^{(1)}$ or $g_{+}^{(2)}$ is irrelevant,
being reabsorbed in $\rho_s$, the {\it relative} correction matters and it is this the relevant quantity to study ---
together with the decay width of $g_{+}^{(2)}$ --- for quantifying the effect
of the resonance. They are also the relevant quantities for the $b_-^{(1)}$ annihilation, once the relation between $\rho_s$ and $k_b$ is fixed.
We have then computed the mass splitting $\Delta m_g \equiv 2 m_{g_-}^{(1)}-m_{g_+}^{(2)}$
at one--loop level. Details on such a computation can be found in the Appendix B.
For the parameter range taken above, the result of the splitting is the following:
\be
\Delta m_{g}=m_{g_{+}}^{(2)}-2m_{g_{-}}^{(1)} \simeq-1.4\, \alpha_{s}\, m_{g_{+}}^{(2)}\,\,\,,
\label{DeltaMg}
\ee
where $\alpha_s$ is the strong coupling constant, evaluated at the energy scale $\rho_s/R$.
The value (\ref{DeltaMg}) is comparable with the total decay rate $\Gamma_g$ of $g_{+}^{(2)}$, which at tree--level
is purely given by the processes $g_{+}^{(2)}\into \bar{q}_{L,R}\,q_{L,R}$.
For each quark, we get $\Gamma_{g,L/R} = \frac{1}{12}\,(c_{L/R,g}^{(2,0,0})^2 \alpha_s m_{g_{+}}^{(2)}$,
where the couplings $c_{L/R,g}^{(2,0,0)}$ are given by Eqs.(\ref{cRres})
and (\ref{cLres}).
Summing over all SM quarks:
\be
\Gamma_{g}= \tilde{c}^2 \alpha_s m_{g_{+}}^{(2)} \simeq 1.5 \, \alpha_s \, m_{g_{+}}^{(2)}\,,
\label{DW}
\ee
where $\tilde{c}^2$ is the mean value of the couplings $c_{L/R,g}^{(2,0,0)}$ squared.
As can be seen, $\Gamma_g\simeq |\Delta m_g|$.

\section{Relic Density}

The setup we have introduced is typical for frameworks embedding a cold dark matter
candidate. There is a tower of massive states which are in thermal equilibrium in the
early Universe, and a symmetry, the $\Z_2$-parity, preventing the lightest of these states
to decay. We have also shown that it is natural for such stable species to be the $A_-^{(1)}$,
i.e. a particle which is electric- and color-charge neutral and hence, potentially, a good
dark matter candidate. As a rule of thumb, the thermal relic density of a massive particle
(i.e. a particle non-relativistic at freeze-out) scales with the inverse  of its pair annihilation
rate into lighter SM species. In case of the $A_-^{(1)}$, we need to take into account that its
mass splitting with other antiperiodic states can be small: there is a full set of "coannihilating"
particles, whose density evolution needs to be described simultaneously through a set
of coupled Boltzmann equations~\cite{Binetruy:1983jf,Griest:1990kh,Edsjo:1997bg}.
The picture is analogous to what one finds for UED models~\cite{Kong:2005hn,Burnell:2005hm},
or sometimes encounters in the supersymmetric frameworks, see
e.g.~\cite{Edsjo:1997bg,Ellis:1998kh,Edsjo:2003us}:
if the annihilation rate per degree of freedom of the slightly heavier states is larger (smaller)
than for the lightest particle, coannihilations tend to delay (anticipate) the decoupling of the latter from the
thermal bath, and hence to diminish (enhance) the thermal relic component. In practice, in
the early Universe thermal environment, coannihilating states are essentially indistinguishable,
since one species is turned into another by inelastic scatterings over background particles
(these interactions tend to be much faster than pair annihilation processes because they are
triggered by relativistic background states). The Boltzmann equation governing the freeze out
is then conveniently rewritten in terms of the total number density $n = \sum_i n_i$, with the
sum running over all coannihilating particles; it takes the form~\cite{Edsjo:1997bg}:
\begin{equation} \label{eq:Boltzmann2}
  \frac{dn}{dt} =
  -3Hn - \langle \sigma_{\rm{eff}} v \rangle
  \left( n^2 - n_{\rm{eq}}^2 \right)\,,
\end{equation}
where $H$ is the Hubble parameter, while $n_{\rm eq}$ is the total equilibrium number density,
i.e., in the Maxwell-Boltzmann regime:
\begin{equation} \label{eq:neq}
  n_{\rm eq} =
  \frac{T}{2\pi^2} \sum_i g_i m_{i}^2
  \, K_{2} \!\left( \frac{m_{i}}{T}\right).
\end{equation}
The effective thermally-averaged annihilation cross section $\langle \sigma_{\rm eff} v \rangle$
drives the decoupling and reads:
\begin{equation} \label{eq:sigmaveff}
  \langle \sigma_{\rm{eff}} v \rangle =
  \frac{1}{n_{\rm{eq}}^2} \, \frac{g_1^2 T}{4 \pi^4} \int_{0}^\infty dp_{\rm eff}
  p^2_{\rm eff}  \, K_{1} \!  \left( \frac{\sqrt{s}}{T}\right)
  \, W_{\rm eff} \left( s\right)\,,
\end{equation}
with all relevant pair-annihilation channels included in the effective annihilation rate:
\begin{equation} \label{eq:rateeff}
   W_{\rm eff} \left( s\right) =  \sum_{ij} \sqrt{\frac{[s-(m_{i}-m_{j})^2][s-(m_{i}+m_{j})^2]}
  {s(s-4m_1^2)}} \frac{g_ig_j}{g_1^2} W_{ij}  \,.
\end{equation}
In the expressions above, $K_{l}(x)$ is the modified Bessel functions of the second kind of
order $l$; $m_{i}$ and $g_i$ are, respectively, the mass and the number
of internal degrees of freedom for the coannihilating particle $i$, with the index
$i=1$ referring to the lightest state.  For all pair annihilation processes the kinematics
has been written in terms of $p_{\rm{eff}}$ and $s = 4 (p_{\rm{eff}}^2 + m_1^2)$,
the center-of-mass momentum and energy squared in the annihilation of a pair of lightest
states; the annihilation process with given initial states $i$ and $j$ needs to be included
in the effective annihilation rate whenever $s \ge (m_i + m_j)^2$.

Relic abundances are computed solving numerically the density evolution
equation~(\ref{eq:Boltzmann2}) with the techniques developed in~\cite{Gondolo:1990dk}
and implemented in the \ds\ package~\cite{Gondolo:2004sc}.
The first step is to derive the expression for $W_{\rm eff} \left( s\right)$ for generic couplings
and mass spectrum in the model. For any given set of the free parameters,
$W_{\rm eff}$ is then provided as input in the \ds\ code which makes a tabulation as a
function of the effective momentum $p_{\rm{eff}}$, taking care of resonances and
coannihilation thresholds. The Boltzmann equation is then integrated numerically in the
variable $Y=n/s$, where $s$ is the Universe entropy density;
thermal equilibrium $Y=Y_{\rm{eq}}$ is assumed as boundary condition at the temperature
$T = m_1/2$, and the evolution is followed up to the point, after freeze-out, when $Y$ settles
on a constant value: the relic density is obtained just by scaling this value to the value
$s_0$ of the entropy density in the Universe today.
Contrary to most analyses in the literature, we do not perform the computation of the relic density
by replacing the thermally averaged annihilation cross--section with a truncated expansion in
powers of $T/m_1$; our procedure gives a more accurate result, especially in case of
coannihilation and resonance effects.

\subsection{Minimal DM framework}
\label{mindm}

We consider first the framework in which the mirror symmetry does not act on
the colored $SU(3)_s$ group, and all gluons are periodic states on $S^1$. In this case,
for the typical set of fermionic parameters introduced in Section~2, the DM candidate
$A_-^{(1)}$ shares large coannihilation effects with the lightest antiperiodic fermion $b_-^{(1)}$ (see Fig.1);
the latter are actually two degenerate fermions in the ${\bf 6}$ of $SU(3)_w$ (see Table~\ref{tabdf} for an account
of the degrees of freedom of $b_-^{(1)}$ and other relevant particles). The antiperiodic fermions
$c_-^{(1)}$ and $\tau_-^{(1)}$ will also be included in the numerical computation
of the relic density, although their contribution is very small. As in all coannihilation
schemes, our results will be most sensitive to the relative mass splitting between the DM
candidate and the heavier state. In what follows we treat the mass of $A_-^{(1)}$ as a free
parameter, or, having fixed the Lorentz violating parameter $\rho =1$, use it interchangeably
with the compactification scale $1/R$,  (recall that  $m_{A_-} = \rho/(2\,R)$). We also take
the mass of $b_-^{(1)}$ as a free parameter;  this is equivalent to introducing a slight departure
of the Lorentz breaking parameter $k_b$ from its unbroken value $k_b=1$, having
assigned  $\epsilon^b_{1,2} = 0.2$,  $k_t = 2.5$ and $\epsilon^t_{1,2} = 1.2$;
for all other antiperiodic fermions we assume $k_i = k_b$ and $\epsilon_{1,2}^i = 0.1$.

\begin{figure}[t]
 \begin{minipage}[htb]{7cm}
   \centering
   \includegraphics[width=6.8cm]{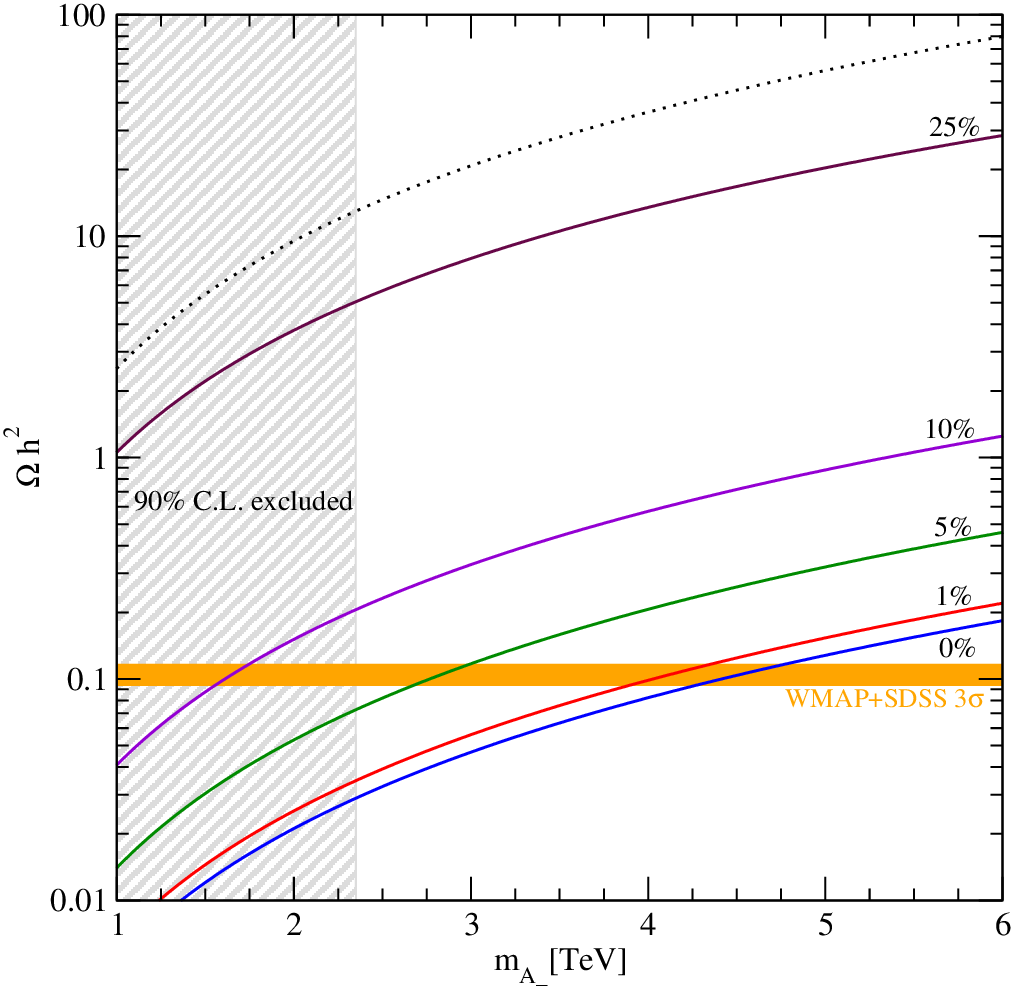}
   \caption{Relic density versus the $A_-^{(1)}$ mass, for a few value of the relative
   mass splitting between $A_-^{(1)}$ and $b_-^{(1)}$, and in the case of the $g_+^{(2)}$
   on resonance in $b_-^{(1)}$ pair annihilations.}
   \label{fig2}
 \end{minipage}
 \ \hspace{2mm} \hspace{3mm} \
 \begin{minipage}[htb]{7cm}
  \centering
   \includegraphics[width=6.8cm]{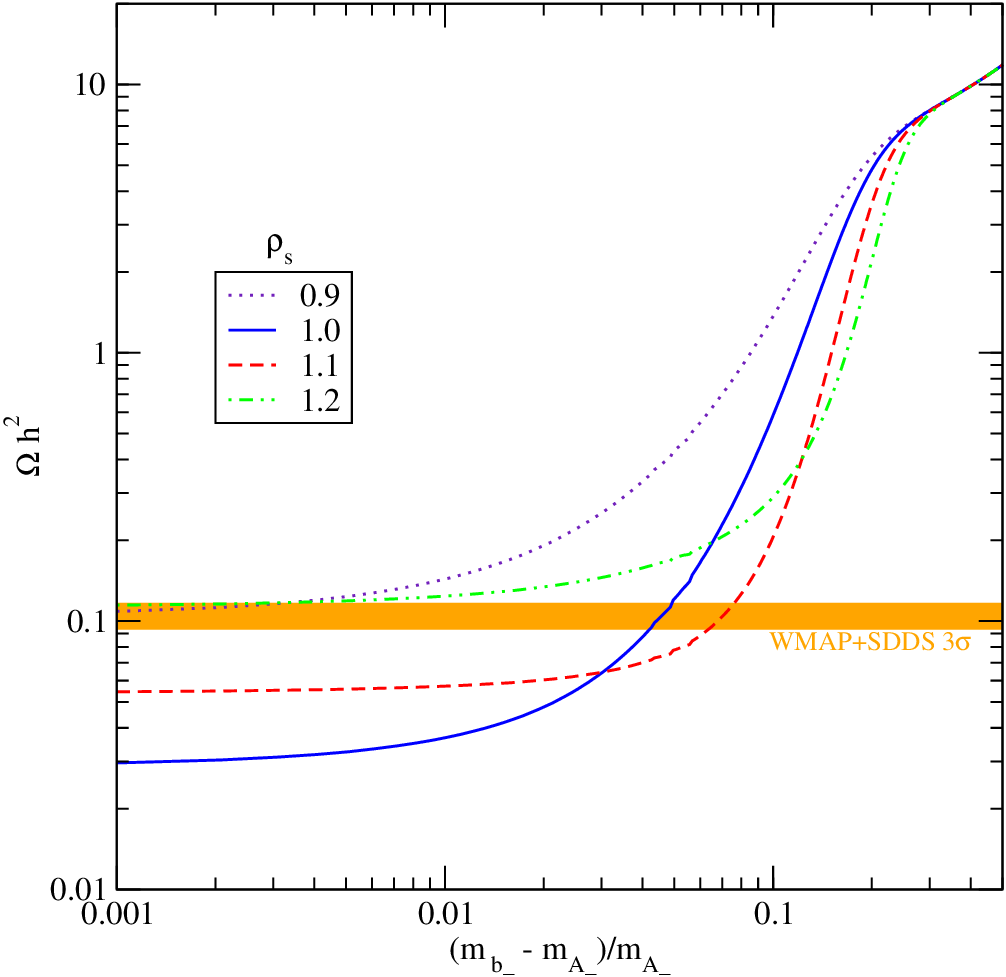}
   \caption{Relic density versus the mass splitting between $A_-^{(1)}$ and $b_-^{(1)}$ for
   a few values of Lorentz breaking parameter $\rho_s$ and assuming as compatification
   scale $1/R=4.7$~TeV.}
   \label{fig3}
 \end{minipage}
\end{figure}

Since electroweak precision tests set a lower bound on the compactification scale
at about $1/R > 4.7$~TeV (90\% C.L., see~\cite{Panico:2006em}), the attempt here
is to introduce a dark matter candidate with a mass of  2.3~TeV or larger. Moreover,
$A_-^{(1)}$ does not minimally couple to the localized fermions, which are the main components
of SM fields. The only diagrams giving a significant contribution to the
$A_-^{(1)}$ pair annihilation rate are those with a third generation quark in the final state
and a third generation antiperiodic fermion in a t- or u-channel; this follows from the fact that
only for the third generation the antiperiodic fermion and gauge boson
wavefunctions can have a order one overlap with SM fields. We list the set of Feynman rules relevant
for this process and the others introduced below in Appendix A, while the full list of the diagrams
which are needed in the relic density calculation is given in Appendix~C.
In the region of interest for our model, we find that, whenever $b_-^{(1)}$ coannihilations
are not effective, the thermal relic abundance of $A_-^{(1)}$ greatly exceeds the cosmological
limit, see the dotted curve in Fig.~\ref{fig2}.

On the other hand, pair annihilation rates for the $b_-^{(1)}$ are much larger and do enter
critically in the effective thermally averaged cross section:
there is a full set of strongly interacting final
states mediated, in the s-channel, by either the SM gluon or by the first periodic KK-gluon
$g_+^{(2)}$. In general, it is not relevant to include in our computation states with KK number
greater than 1; in this case, however, since $m_{g_+}^{(2)}$ is about twice $m_{b_-}^{(1)}$
(recall that $m_{g_+}^{(2)}$ is of order $1/R$, while $m_{b_-}^{(1)}$ of order $1/(2\,R)$),
the annihilation diagrams with $g_+^{(2)}$ in the s-channel become resonant and tend to
give the dominant contribution to the cross section (the effect of resonances induced
by second KK particles was first pointed out in~\cite{Kakizaki:2005en} within the UED context).
The enhancement is maximized for splittings
$\left|2 m_{b_-}^{(1)} - m_{g_+}^{(2)}\right|$ which are below the  $g_+^{(2)}$ decay width,
see Eq.~(\ref{DW}), which is indeed much smaller than the energy flowing in the s-channel.
We find that, on top of the two mass
parameters $m_{A_-}^{(1)}$ and $m_{b_-}^{(1)}$, the mass of  $g_+^{(2)}$ is the third
unknown entering critically in our analysis; we take it as a free parameter,
again in connection to a possible mild variation of $\rho_s$  around its non-violating Lorentz
value $\rho_s =1$.

Finally, there is another relevant issue concerning strongly interacting states we wish to
mention before going to the illustration of results:
we are considering processes taking place at a center of mass energy $\simeq 1/R$ which
is about twice the mass of the annihilating DM particle. The QCD coupling constant $\alpha_s$ should
be evaluated at this relatively high scale and hence renormalization group effects cannot be neglected,
in principle. Indeed, the DM abundance is highly sensitive to $\alpha_s$ which
enters quadratically in annihilation rates: roughly $\Omega_{DM}\propto \alpha_s^{-2}$.
We have computed the one--loop $\beta$-function for
$\alpha_s$ within our framework (see Appendix D for details) and implemented the running numerically in our
Boltzmann code; at 5 TeV, $\alpha_s$ turns out to be approximately 0.097.\footnote{The
running of  $\alpha_s$ was apparently overlooked in
Refs.~\cite{Kong:2005hn,Burnell:2005hm} when estimating the effect of coannihilations of
the LKP with strongly interacting KK states within the UED framework. As explained in Appendix~D,
the effect in UED is larger than for our model; since annihilation cross sections scale
approximately as  $\alpha_s^2 \times (1/R)^{-2}$, we expect that the values of $1/R$
inferred from the cosmological limit should be correspondingly rescaled down.}

\begin{figure}[t]
\centering
\includegraphics[width=11cm]{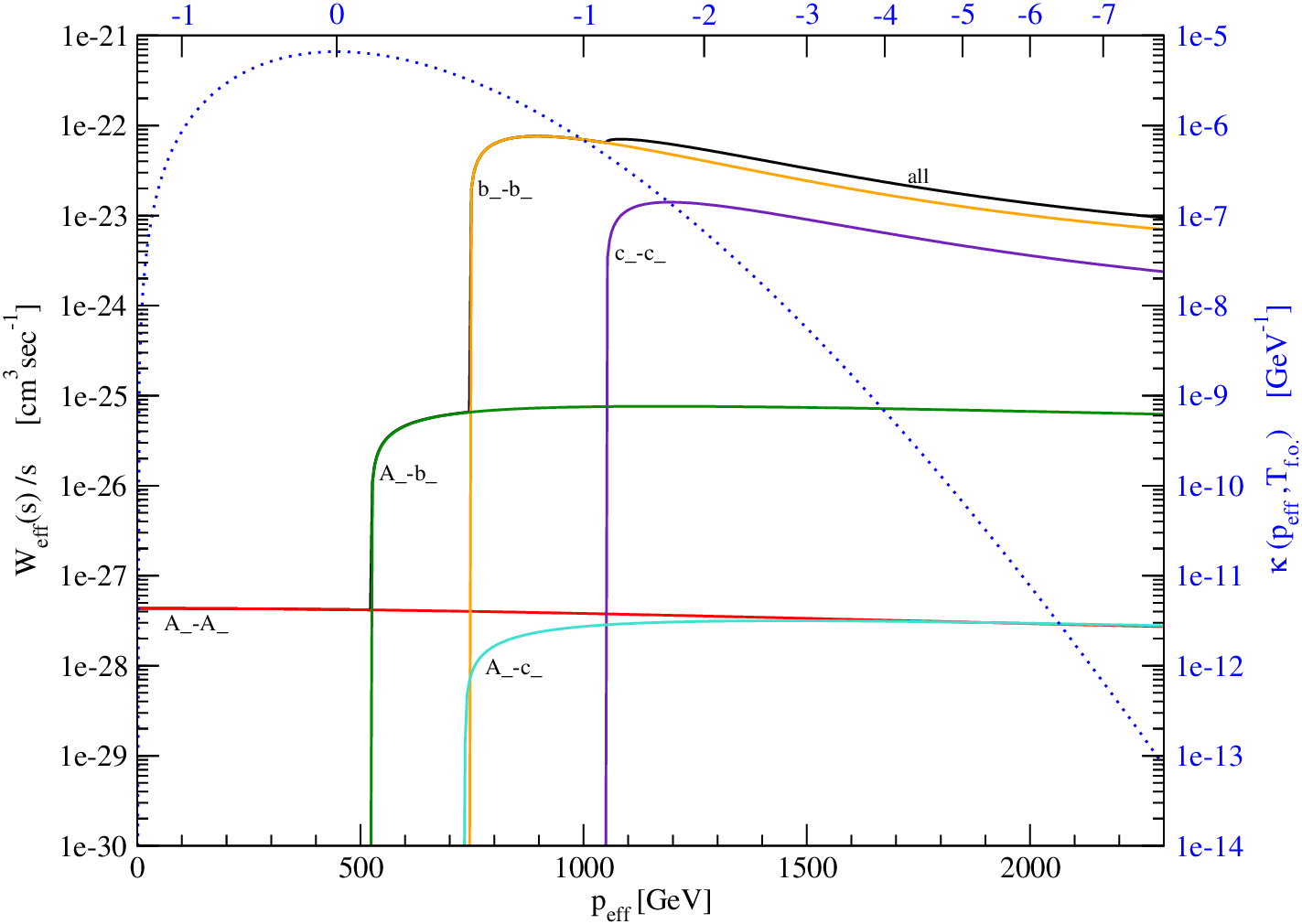}
\caption{Effective annihilation rate $W_{\rm eff}$ over the center of mass energy squared
$s$, plotted versus the effective center of mass momentum $p_{\rm{eff}}$. Contributions from
single annihilation and coannihilation channels are displayed. Also shown (dotted line) is the
thermal weight function $\kappa$ (units of GeV$^{-1}$ as displayed with the scale on the
right-hand side of the plot).}
\label{fig4}
\end{figure}

In Fig.~\ref{fig2} we show the results for the relic density as a function of $m_{A_-}^{(1)}$, for a
few values of the relative mass splitting $(m_{b_-}^{(1)} - m_{A_-}^{(1)})/m_{A_-}^{(1)}$, and taking
$g_+^{(2)}$ on resonance, i.e.  $m_{g_+}^{(2)} \equiv 2 m_{b_-}^{(1)}$. From the case with zero
mass splitting one can read out the cosmological upper limit on $m_{A_-}^{(1)}$ within this
framework, namely $m_{A_-}^{(1)} \leq 4.5$~TeV, or equivalently the bound on the compactification scale
$1/R \leq 9$~TeV: this scale is comparable to those favoured by electroweak precision tests
\cite{Panico:2006em}.
As expected, the prediction of the relic density scales rather rapidly to larger values when
the mass splitting among the coannihilating states is increased, and, consequently, the
value for the mass of the DM candidate approaches the region excluded by electroweak physics
(in the figure, the light-shaded horizontal band correspond to the 3~$\sigma$ preferred region
from the combined analysis of  data on the CMB from WMAP and from the SDSS large scale structure
survey~\cite{Tegmark:2006az}; models which lay below the band correspond to configurations
in which $A_-^{(1)}$ accounts for only a portion of the DM in the Universe, while those above
it are cosmologically excluded). In Fig.~\ref{fig3} we plot the thermal relic abundance as
a function of the mass splitting of the coannihilating states for a model with the minimum
allowed compatification radius $1/R = 4.7$~TeV, and for a few values of $\rho_s$.
In this case, as it can be seen, cosmological constraints restrict the Lorentz breaking parameter
of $SU(3)_s$ roughly in the range $[0.9,\,1.2]$; the interval is not symmetric around
$\rho_s =1$ since, in the Boltzmann equation, annihilations take place at a finite temperature.
For $\rho_s <1$, temperature corrections drive the process at energies always slightly
above the resonance. In the opposite regime the resonance is always met, provided one
considers sufficiently high temperatures; if $\rho_s$ is large, the temperature at which the
resonance is hit is too large compared to the freeze-out temperature and the models
becomes cosmologically excluded. Curves for the four sample values of $\rho_s$ overlap
at a mass splitting  of about 30\%, beyond which coannihilation effects induce negligible changes
on the $A_-^{(1)}$ relic abundance.

\begin{figure}
 \begin{minipage}[b]{7cm}
   \centering
   \includegraphics[width=6.8cm]{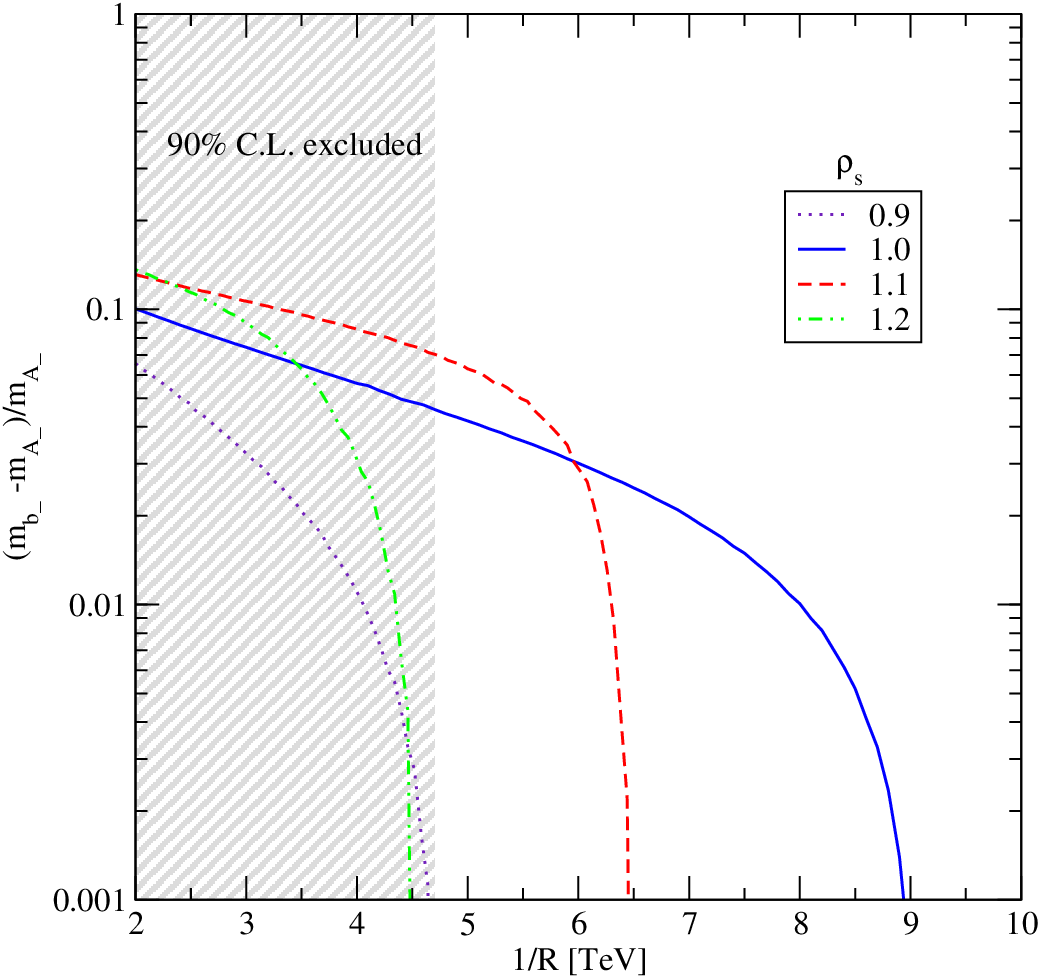}
   \caption{Correlation between the $A_-^{(1)} - b_-^{(1)}$ mass splitting and compactification radius
   in models with $\Omega h^2$ matching the best fit value from cosmological observations,
   for a  few values of $\rho_s$.}
   \label{fig5}
 \end{minipage}
 \ \hspace{2mm} \hspace{3mm} \
 \begin{minipage}[b]{7cm}
  \centering
   \includegraphics[width=6.8cm]{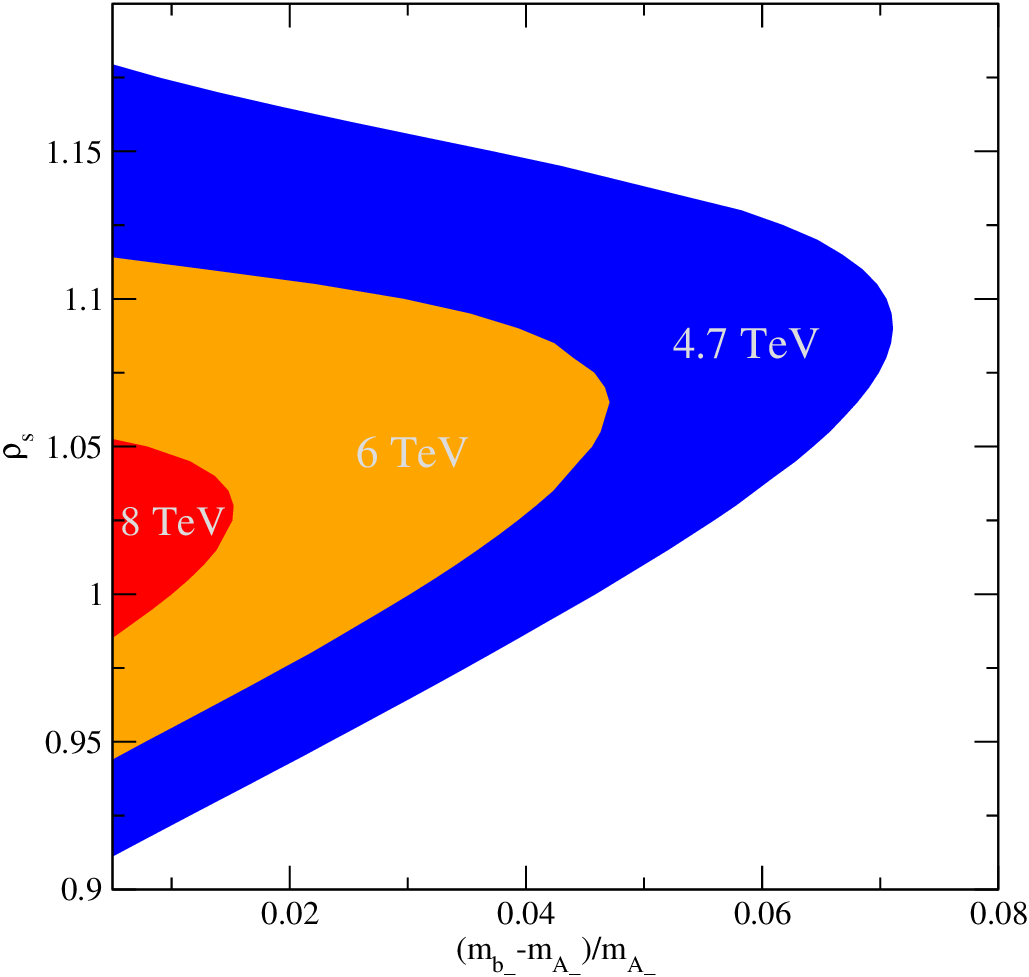}
   \caption{For a few selected values of the compactification scale, regions  in the parameter space
   $\rho_s$--mass splitting in which $\Omega h^2$ is lower or equal to the best fit value from
   cosmological observations.}
   \label{fig6}
 \end{minipage}
\end{figure}

Conservatively, we include in the relic density calculation all states with a mass splitting
below 50\%. In Fig.~\ref{fig4} we illustrate better the role of coannihilations and of the
Boltzmann suppression when mass splittings become too large. We consider the model with
$1/R=4.7$ TeV, $\rho_s=1$ and $k_i =1$, with relic density of about $0.1$,
and plot the effective annihilation rate
$W_{\rm eff}$, as defined in Eq.~(\ref{eq:rateeff}), over the center of mass energy squared
$s$, as a function of the effective center of mass momentum $p_{\rm{eff}}$. Contributions to
$W_{\rm eff}$ from the individual annihilation and coannihilation processes are shown;
coannihilations appear here as thresholds at $\sqrt{s}$ equal to the sum of the
masses of the coannihilating particles, with the NLKP entering first, and $c_-^{(1)}$
at a slightly larger $p_{\rm{eff}}$. The threshold effects are so sharp since coannihilation rates
are large, but also because the number of internal degrees of freedom for the antiperiodic fermions
is much larger than that for $A_-^{(1)}$ (see Table~\ref{tabdf}).
Also shown in Fig.~\ref{fig4} is the weight function
$\kappa(p_{\rm{eff}},T)$ defined implicitly by rewriting the thermally averaged cross section in
Eq.~(\ref{eq:sigmaveff}) as
\begin{equation} \label{eq:sigmaveff2}
  \langle \sigma_{\rm{eff}}v \rangle \equiv \int_0^\infty
\!\!  dp_{\rm{eff}} \frac{W_{\rm eff} \left( s\right)}{s}
 \, \kappa(p_{\rm{eff}},T)\;.
\end{equation}
The function $\kappa$ contains the Boltzmann factors (hence it is exponentially suppressed at large $p_{\rm{eff}}$)
and a phase-space integrand term (hence it goes to zero in the $p_{\rm{eff}} \rightarrow 0$ limit).
It can be view as a weight function, since at any given temperature $T$, it selects the range of
$p_{\rm{eff}}$ which are relevant for the thermal average. In Fig.~\ref{fig4}, $\kappa$ is plotted as
a function of $p_{\rm{eff}}$ at the freeze-out temperature (defined as the temperature
at which $Y$ is equal to twice the final asymptotic value) in units of GeV$^{-1}$ and with
respect to the scale shown on the right-hand side of the plot. On the top of the panel, the tick mark
with the label '0' corresponds to the momentum at which $\kappa$ has its maximum,
while the tick mark with label $-n$ indicates the momentum at which $\kappa$ is $10^{-n}$ of its
peak value. Coannihilation effects are relevant if they provide a significant  enhancement
in the effective annihilation rate within the range of momenta in which $\kappa$ is not too small compared to its peak value; this is clearly the case for the $b_-^{(1)}$ in the example displayed.
Also notice that the effect induced by the $c_-^{(1)}$ is not negligible by itself, however it gets
marginal when superimposed to the one from the $b_-^{(1)}$.

Figs.~\ref{fig5} and \ref{fig6} summarize the picture within our minimal DM framework.
We select models whose thermal relic density matches the best fit value from cosmological
observations $\Omega_{DM} h^2 = 0.105$. As already explained, there are three relevant mass
parameters in the model: $m_{A_-}^{(1)}$ or equivalently $1/R$, $m_{b_-}^{(1)}$ or equivalently
the relative mass splitting between $b_-^{(1)}$ and $A_-^{(1)}$, and $m_{g_+}^{(2)}$ or
equivalently $\rho_s$. In Fig.~\ref{fig5} we select a few values of $\rho_s$ and find the isolevel curves
for $\Omega h^2$ in the plane of the other two; in Fig.~\ref{fig6}, instead, a few values
of the compatification scale are considered and correlations between the other two parameters
derived. The general trends we see here are essentially along the lines we have already discussed for
Figs.~\ref{fig2} and \ref{fig3}; we display more clearly the strict upper limits on $1/R$ (about 9~TeV),
and find that the NLKP--LKP mass-splitting needs to be at about the 7\% level or smaller.
The required range of $\rho_s$, for a given compactification scale and mass splitting, is also
displayed.

We have definitely found a tight interplay among the parameters in the model;
the procedure of embedding a DM candidate in this minimal scenario has been successful,
pointing to a limited set of patterns in the parameter space.

\subsection{DM in the framework with a copy of $SU(3)_s$}

If the $\Z_2$ mirror symmetry acts on the colored $SU(3)_s$ group,
the first antiperiodic KK gluon $g_-^{(1)}$, which has a mass of order $1/(2\,R)$, enters
critically in the computation of the relic abundance for the $A_-^{(1)}$. In most extensions to the SM,
strongly interacting gauge bosons are the particles with largest pair annihilation cross section
per internal degree of freedom, hence tend to give the largest possible coannihilation effects.
This has been verified also in the extra-dimension context studying the coannihilation of the LKP
with the first KK excitation of the gluon in UED~\cite{Kong:2005hn,Burnell:2005hm}.

\FIGURE[t]{
\centerline{
\epsfig{file=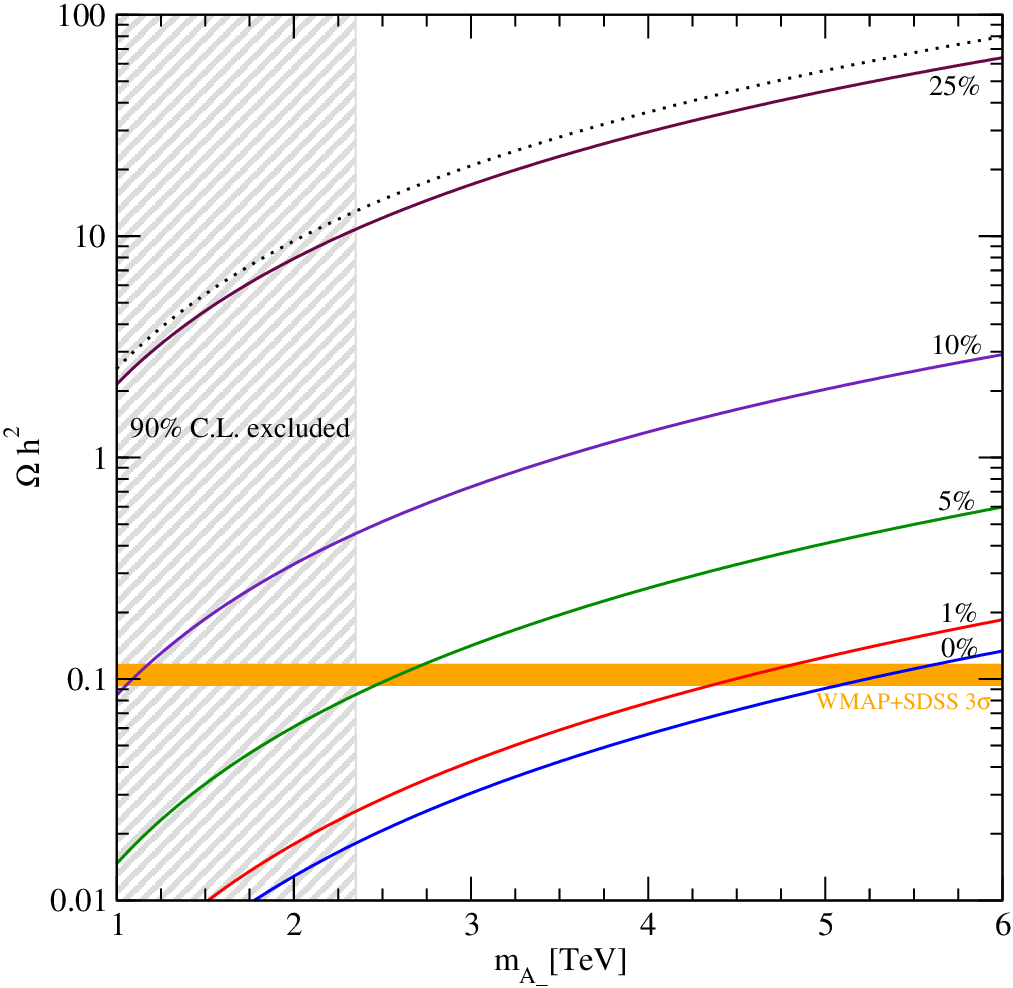,width=7.cm}\quad $\;\;$\quad
\epsfig{file=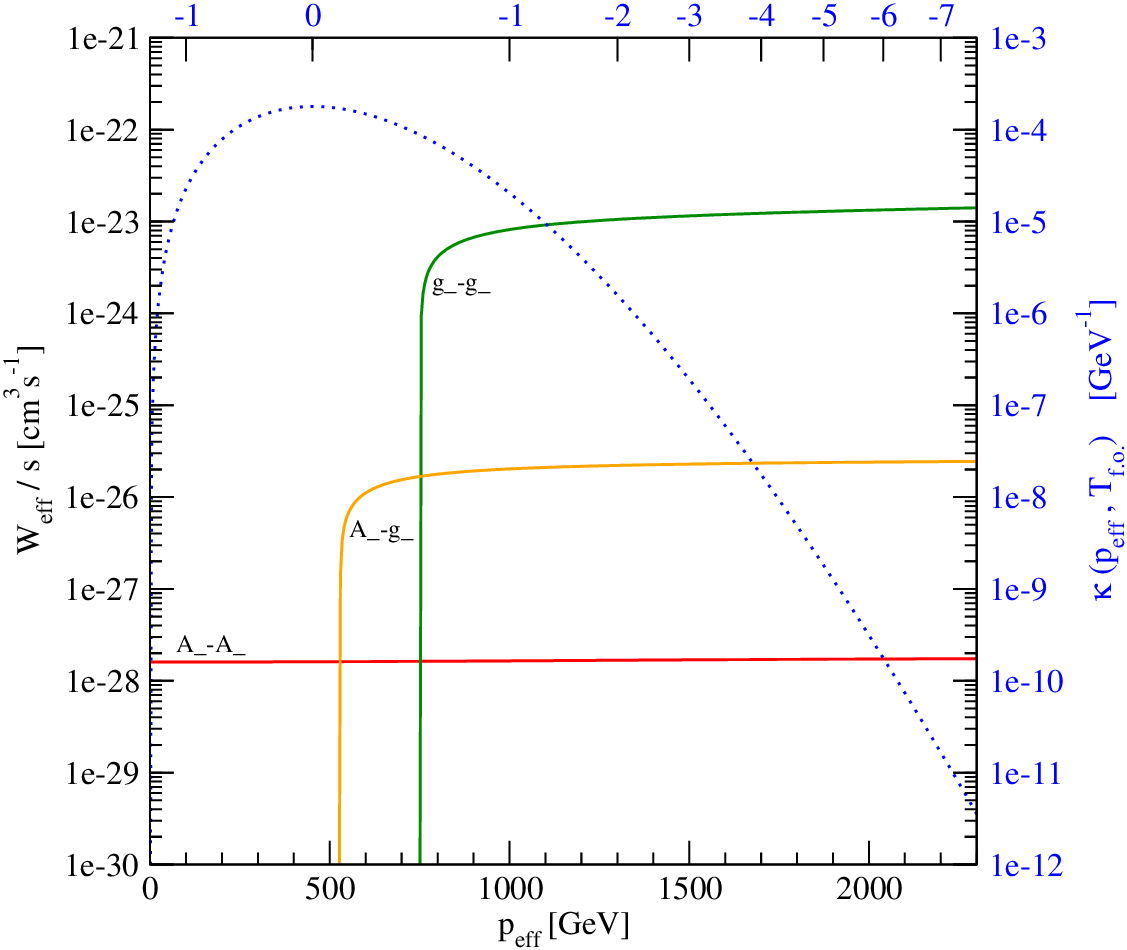,width=8.2cm}}
\caption{{\it Left Panel:} Relic density versus the $A_-^{(1)}$ mass, for a few value of the relative
   mass splitting between $g_-^{(1)}$ and $A_-^{(1)}$. Antiperiodic fermions have been
   decoupled assuming they have a mass splitting larger than 50\%.
  \textit{Right Panel:} Effective annihilation rate $W_{\rm eff}$ over the center of mass energy squared
   $s$, as in Fig.~\protect{\ref{fig4}}, but for a model with $g_-^{(1)}$ coannihilations decreasing
   the relic density of $A_-^{(1)}$ to the level of the best fit from cosmological observations.
   The thermal weight function $\kappa$ is shown as a dotted curve; see Fig.~\ref{fig4}, and the
   relative  discussion in the text, for further details.}
 \label{fig7}}

We discuss the phenomenology in our model referring again to the three mass parameters
introduced above. Note, however, that in this case we select values of $\rho_s$ to fix both
the mass of $g_-^{(1)}$ (we implement the tree-level relation $m_{g_-}^{(1)} = \rho_s/(2\,R)$) and
the mass of $g_+^{(2)}$ (through the 1-loop mass splitting as found in Eq.~(\ref{DeltaMg})).
We start by examining the effect of $g_-^{(1)}$ coannihilations alone. In Fig.~\ref{fig7}
we set $k_b = k_i=1.5$, removing all antiperiodic fermions from the coannihilation list,
and discuss the effect of degeneracies in mass between $g_-^{(1)}$ and $A_-^{(1)}$.
In the limit of zero mass splitting we find as upper bound on the compactification scale
$1/R \leq 11$~TeV. As expected, the bound on $1/R$ found within the minimal scenario
has been relaxed. We also find that, at the lowest allowed value for $1/R$,
$(m_{g_-}^{(1)}-m_{A_-}^{(1)})/m_{A_-}^{(1)}\leq 6\%$ must old.  Even in the present framework,
$g_-^{(1)}$ coannihilations appear as sharp thresholds in the invariant rate. The channels
contributing to the annihilation rate are listed in Appendix~C. They include the case of annihilation into
quarks with the $g_+^{(2)}$ in a s-channel; however, this process always takes place
slightly off--resonance, since $|\Delta m_g|$ is of the same size as  $\Gamma_g$, and it is then
always subdominant with respect to the process with gluon final states
(one may compare the behavior  of the $g_-^{(1)}$-$g_-^{(1)}$ term in the right panel of
Fig.~\ref{fig7} as a function of  $p_{\rm{eff}}$, with the analogous for the $b_-^{(1)}$-$b_-^{(1)}$
term in Fig.~\ref{fig4}, where the enhancement due to the resonance is instead evident).

\begin{figure}
 \begin{minipage}[b]{7cm}
   \centering
   \includegraphics[width=6.8cm]{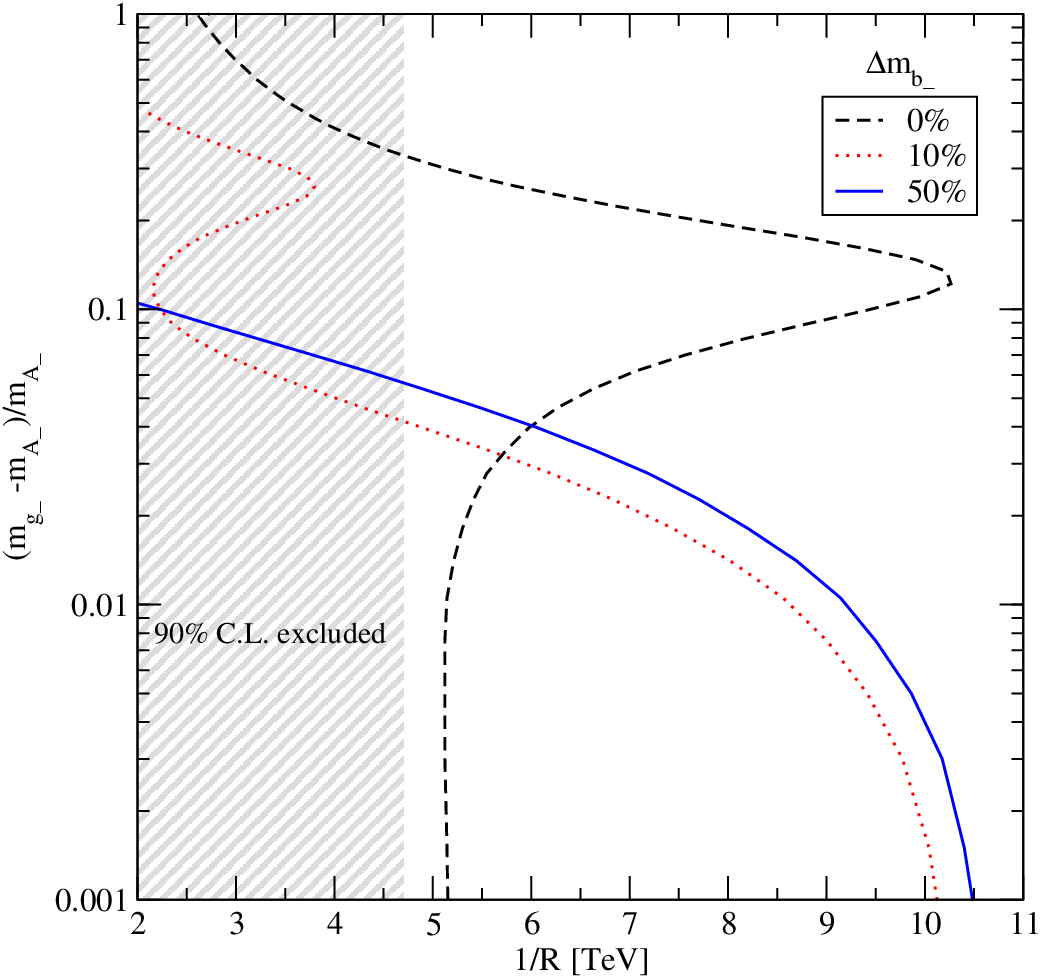}
   \caption{Correlation between the $A_-^{(1)} - g_-^{(1)}$ mass splitting and
   compactification scale for models with $\Omega h^2$ matching the best fit value from
   cosmological observations, and for a few values of $A_-^{(1)} - b_-^{(1)}$
   mass splitting.}
   \label{fig9}
 \end{minipage}
 \ \hspace{2mm} \hspace{3mm} \
 \begin{minipage}[b]{7cm}
  \centering
   \includegraphics[width=6.8cm]{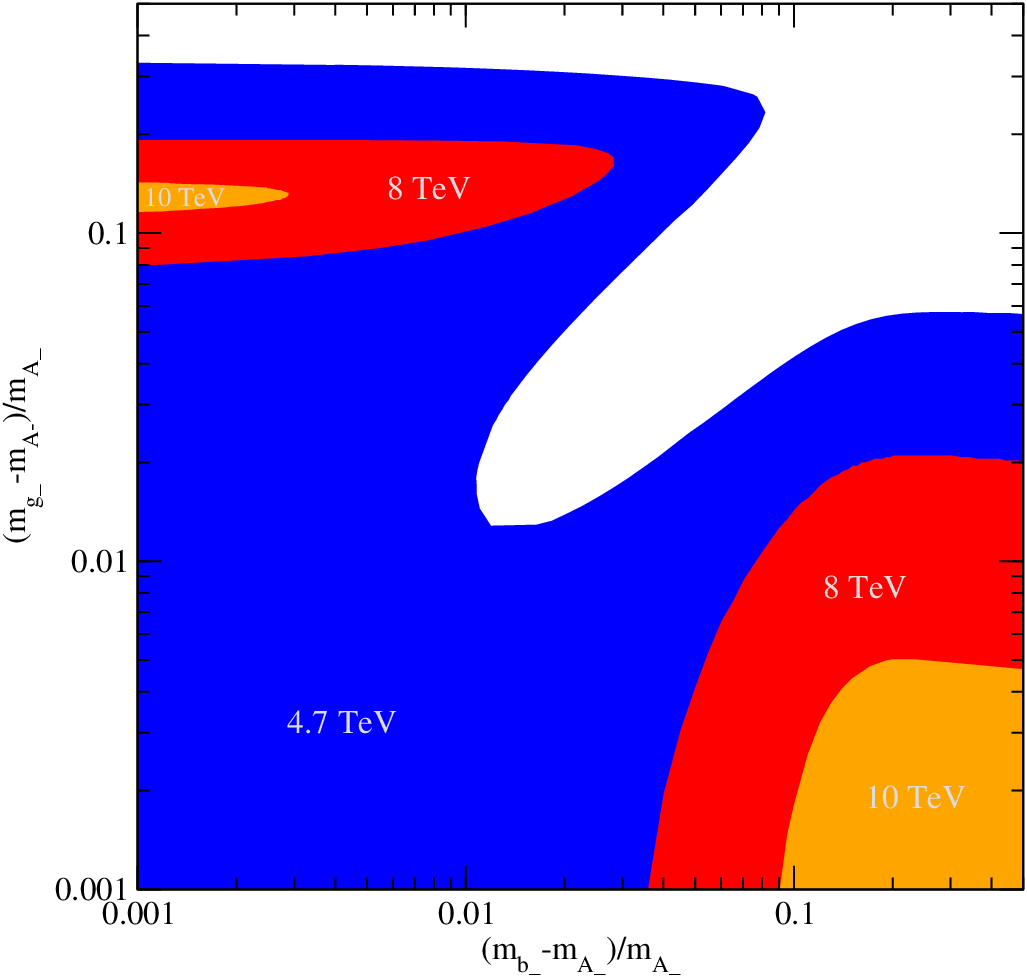}
   \caption{For a few selected values of the compactification scale, regions  in the parameter space
    $m_{g_-}^{(1)}-m_{A_-}^{(1)}$ versus $m_{b_-}^{(1)}-m_{A_-}^{(1)}$ in which $\Omega h^2$ is
    lower or equal to the best fit value from cosmological observations.}
   \label{fig10}
 \end{minipage}
\end{figure}

The general framework, with both $g_-^{(1)}$ and $b_-^{(1)}$ playing a role in the relic density
calculation, is illustrated in  Figs.~\ref{fig9} and \ref{fig10}. The picture is not a mere overlap of
two distinct coannihilation effects. As we have already mentioned, at a given $1/R$ and given
mass splitting between $b_-^{(1)}$ and $A_-^{(1)}$, the mass splitting between $g_-^{(1)}$ and
$A_-^{(1)}$  sets also $m_{g_+}^{(2)}$ and hence whether the $b_-^{(1)}$ pair annihilation is
resonantly enhanced or not. The second effect is due to the fact that we are
superimposing coannihilations from states with different annihilation strength, and, especially,
different number of internal degrees of freedom ($g=24$ for $g_-^{(1)}$):
for equal mass splitting, the matching needs to be done at the level of annihilation rates per
degree of freedom. The net effect can be both of increasing or lowering the thermal relic abundance
for $A_-^{(1)}$. E.g., if we add, on top of a configuration with efficient $g_-^{(1)}$ coannihilations, a
$b_-^{(1)}$ state with small mass splitting with the $A_-^{(1)}$, but with mass significantly
displaced from the $g_+^{(2)}$ resonance, we are effectively including a set of passive
degrees of freedom: maintaining the tower of states in thermal equilibrium becomes more
energetically  expensive, the freeze-out is anticipated and the thermal relic density increased.
This is what happens at small  $m_{g_-}^{(1)}-m_{A_-}^{(1)}$ and small
$m_{b_-}^{(1)}-m_{A_-}^{(1)}$ in the throat region of Fig.~\ref{fig10}.

Introducing the $g_-^{(1)}$ in the framework has enlarged the regions in the parameter
space which are compatible with the cosmological constraints; still, the tight correlation
patterns among parameters in the model, which had emerged in the minimal scheme, are maintained
here, although in slightly different forms.

This feature might  be view as a sign of fine tuning. To better quantify this point,
in analogy to the study of naturalness of radiative electroweak symmetry
breaking~\cite{Barbieri:1987fn}, we introduce the fine-tuning measure~\cite{King:2006tf,Ellis:2001zk}:
\begin{equation}
\displaystyle
\Delta^\Omega\equiv {\rm max}
\left\{\frac{\partial \ln ( \Omega h^2)}{\partial \ln(a)}\right\},
\end{equation}
where $a$ labels any of the free parameters in our model.
In the minimal DM framework of Section~\ref{mindm}, $\Delta^\Omega$ changes from about 35 in the
lower part of Fig.~\ref{fig6} to about 8 for the models with largest $\rho_s$. In the model with antiperiodic gluons,
the parameter  space with small $m_{b_-}^{(1)}-m_{A_-}^{(1)}$ and intermediate $m_{g_-}^{(1)}-m_{A_-}^{(1)}$
in Fig.~\ref{fig10}
has a minimum $\Delta^\Omega$ of about 7, while in the limit  of large $m_{b_-}^{(1)}-m_{A_-}^{(1)}$
a fine-tuning correlated to the ${A_-}^{(1)}$--${g_-}^{(1)}$ mass splitting of about 34; finally in the
throat region, in which both mass splittings are small, the interplay among the parameters reaches its
maximum and, correspondingly,  $\Delta^\Omega$ can be as large as 50.
Such moderate degree of fine-tuning ($\Delta^\Omega \le 10$ is expected in a
``natural" model) is comparable or even smaller than what one obtains in other cases
in the literature when the relic density of the WIMP DM candidate is driven by coannihilation
effects, see, e.g.,~\cite{Ellis:2001zk}.

 \section{Discussion and conclusion}

 We have shown how to embed WIMP dark matter candidates
 into non-universal flat higher dimensional theories aiming at the stabilization of the
 electroweak scale.  We have focussed on the model of \cite{Panico:2006em} and
 shown that, in a large fraction of the parameter space, the lightest antiperiodic particle
 is a stable gauge field with the correct properties for being identified with the DM
 in the Universe. Although electroweak bounds force its mass
 to be larger than about 2.3~TeV, and its interaction rate is rather small, coannihilation
 and resonance effects involving colored particles can delay its decoupling from thermal
 equilibrium, and allow its relic abundance to be within the range currently
 favored by cosmological observations.

 The picture we have introduced is rather unusual,
 since the WIMP dark matter candidate is significantly more massive than in standard
 (e.g. SUSY) scenarios, and its coupling to the SM is essentially limited to third generation
 quarks. The phenomenology of DM searches for this model is less appealing than in
 other frameworks; in particular its scattering rate on ordinary matter is suppressed and
 mediated mainly by radiative effects involving virtual bottom and top quarks. Moreover,
 its zero temperature pair annihilation rate (again mainly into bottom and top quarks)
 is small, at the level of few times $10^{-28}$~cm$^3$~s$^{-1}$ (see Figs.~\ref{fig4}
 and~\ref{fig7}), making it hard to detect annihilation signals in dark matter halos.

  On the other hand, embedding the dark matter candidate in the model of \cite{Panico:2006em}
 introduces favored patterns in the parameter space; tests of this framework at future colliders
 may indeed give crucial information on the DM scenario proposed in this paper.

It would be interesting to implement the mirror symmetry used in this paper to achieve
a DM candidate in warped models as well, mainly for the case of gauge and matter fields in the bulk.
One could for instance double the $U(1)_Y$ gauge field
and identify the DM candidate as the lightest mode of $A_-$. Since couplings to fermions might be larger
and the reference mass scale lower with respect to the flat case, it might be feasible to match the correct relic density without coannihilation effects with colored particles. On the other hand, in order to make
such construction as natural as in our case, it would be desirable to find some other
motivation to introduce the mirror symmetry.

\section*{Acknowledgements}

We would like to thank R. Iengo, G. Panico and G. Servant for useful discussions.
This work is partially supported by the European Community's Human
Potential Programme under contracts MRTN-CT-2004-005104 and MRTN-CT-2006-035863,
and by the Italian MIUR under contract PRIN-2003023852. M.S. and P.U. would like to thank
the Galileo Galilei Institute for Theoretical Physics for hospitality during the preparation of
this work.

\appendix

\section{Feynman Rules}

In this appendix we give some details about the Feynman rules of
our model,  focusing in particular on vertices relevant for the calculations of Section 3.
The Lagrangian (aside from ghosts and gauge--fixing terms)
is given in Eqs.(2.4)--(2.7) of \cite{Panico:2006em}. The gauge--fixing terms
(and the corresponding ghost terms) we use are of the form
\be
{\cal L}_{gf} = -\frac{1}{2\xi} \sum_a\, (\partial_\mu A^{\mu,a} - \xi \, \rho \, \partial_5 A_5^a)^2\,,
\label{gf}
\ee
for all gauge groups. All cross--sections have been evaluated in the $\xi=1$
gauge. Since ghosts and gauge bosons $A_{\mu}$, $A_{5}$ are purely bulk fields,
ghost, 3-bosons and 4-bosons vertices are easily derived from the usual standard ones.
One has only to take into account the Lorentz violation
in the fifth dimension replacing $A_{5}\rightarrow\rho A_{5}$,
$\partial_{5}\rightarrow\rho\partial{}_{5}$ and take the linear combinations
$\phi_{\pm}=(\phi_1\pm \phi_2)/\sqrt{2}$ for $U(1)_i$ and $SU(3)_{i,s}$ gauge and ghost fields.

Fermion-gauge couplings are more involved, due to the non-trivial profile of fermions in the extra
dimensions. The interactions between a gauge boson KK mode $p$
with fermion KK modes $m$ and $n$ can be written as
$iT^{a}g_{4}\gamma^{\mu}(c_{L,a}^{(m,n,p)}P_{L}+c_{R,a}^{(m,n,p)}P_{R})$. The coupling $g_4$
is the 4D gauge coupling, related to the 5D one as $g_{4}= g_{5}/\sqrt{2\pi R}$, the indices
$p,m,n$ run over even (odd) integers for $\Z_2$ even (odd) fields and
$c_{L/R,a}^{(m,n,p)}$ are the integrals of the wavefunctions along the 5th dimension
involving respectively left and right fermion components and broken or unbroken
gauge field components $A_{\mu,a}$.
In terms of the mode expansion
(see Appendix of \cite{Panico:2006em} for further details)
\be
\begin{split}
\Psi_{L/R} =& \sum_{n}f_{L/R}^{(n)}(y)\chi_{L/R}^{(n)} \,, \\
\tilde{\Psi}_{L/R}  =& \sum_{n}\tilde{f}_{L/R}^{(n)}(y)\chi_{L/R}^{(n)}\,,\\
q_{L/R} =& \sum_{n}g_{L/R}^{(n)}\chi_{L/R}^{(n)}\,,
\end{split}
\label{psiDeco}
\ee
where $n$ in Eq.(\ref{psiDeco}) is even (odd) for periodic (antiperiodic) fermions, one gets
\be
\!\!\! c_{L/R,a}^{(m,n,p)}=\sqrt{2\pi R}\int_{0}^{2\pi R}\!\!\!\!\!dy\,f_{\mu,a}^{(p)}(y)
\Big[f_{1,L/R}^{(n)}(y)f_{2,L/R}^{(m)}(y)+\widetilde{f}_{1,L/R}^{(n)}(y)
\widetilde{f}_{2,L/R}^{(m)}(y)+g_{1}^{(n)}g_{2}^{(m)}\delta(y)\Big]\,,
\label{cLR}
\ee
where $f_{\mu,a}^{(p)}(y)$ is the wave--function of the $p^{th}$ KK mode of $A_{\mu,a}(y)$.

As one can check from the Feynman diagrams listed in Appendix C,
the relevant couplings in our calculation are:

$\bullet$ $p=0,m=n$: only gauge bosons of the unbroken SM $SU(2)_L\times U(1)_Y\times SU(3)_s$
gauge group have zero modes, with a constant wavefunction: $f_{\mu ,a}^{(0)}=1/\sqrt{2\pi R}$.
The integral in square brackets in Eq.~(\ref{cLR}) is normalized to be 1 in order to have canonical fermion kinetic
terms:
\be
c_{L/R,a}^{(0,n,n)}=1,
\ee
implying universal couplings for all fermions,
as expected from the unbroken gauge symmetry.

$\bullet$ $p=m=1,n=0$: one gets
\bea
c_{R,a}^{(1,1,0)} & = &
\pm\frac{k \left( k \mp M R \right) \epsilon }
    {{\sqrt{2} \pi R M}\,\left( k^2 + M^2 R^2 \right) \,
      \sqrt{Z_2}} \,, \label{cRexpr} \\
c_{L,a}^{(1,1,0)} & = &
\pm \frac{k_i
      \left( k_i \mp M_i R \right)
      \epsilon_i}{\sqrt{2} \pi M_i R
      \left(k_{i}^2 +
        M_{i}^2 R^2 \right) \,
      \sqrt{Z_1}}\,.
\label{cLexpr}
\eea
In Eqs.(\ref{cRexpr}) and (\ref{cLexpr}), the two different signs refers to the two towers of antiperiodic fermions with same mass and quantum numbers and the $Z$ factors are those appearing in Eq.(2.18) of
\cite{Panico:2005dh} taken at $\alpha=0$ (no EWSB).
These factors are typically $\simeq 1$, aside from the top quark where
they can be substantially bigger ($\simeq 4$ in the chosen setup).
In Eq.~(\ref{cLexpr}), $i=u,d$, depending on the doublet component,
and $M$ in Eq.~(\ref{cRexpr}) should be identified with $M_u$ or $M_d$, depending on the singlet field under consideration.
Similarly for $\epsilon$ and $k$.
Antiperiodic fermion and gauge boson wavefunctions vanish at $y=0$
and thus the overlap with the SM $n=0$ fields is small, $O(\epsilon)$, except for the top and
the left--handed bottom quark, for which one has an overlap $\sim O(1)$.

$\bullet$ $p=2,m=n=0$: we are interested only to the fermion gauge couplings to $g^{(2)}$, the first KK mode
of $SU(3)_s$. One gets
\bea
c_{R,g}^{(2,0,0)} & = & \sqrt{2}
\bigg[1  +
      4\epsilon^2 \frac{MR}{\pi k(k^2+ 4M^2 R^2)} \coth \Big(\frac{\pi M R}{k}\Big)\bigg]Z_2^{-1} \,,
\label{cRres}\\
c_{L,g}^{(2,0,0)} & = &
\sqrt{2}
\bigg[1  + 4 \sum_{i=u,d} \epsilon_i^2 \frac{M_i R}{\pi k_i (k_i^2+ 4M_i^2 R^2)} \coth \Big(\frac{\pi M_i R}{k_i}\Big)\bigg]
Z_1^{-1}\,.
\label{cLres}
\eea
This is a KK-number violating coupling, due to the localized Lagrangian term.
As can be seen from Eqs.(\ref{cRres}) and (\ref{cLres}), this coupling is $\sim \sqrt{2}$ for all SM fermions, but the top and
the left--handed bottom for which it is much smaller ($\sim \sqrt{2}/Z_2^{t}$ ).

$\bullet$ $p=2,m=n=1$: again, the only coupling relevant for us is the one with $g^{(2)}$. Only bulk fields
are involved and the computation is trivial, giving
\be
c_{L/R,g}^{(2,1,1)}=\frac{1}{\sqrt{2}}.
\ee

All effects involving KK states with $p\geq 2$, with the exception of the possible gluon resonance state for $p=2$, have been neglected.

Analogous considerations can be done for the couplings between fermions and the would--be Goldstone bosons $A_{5}$.
The vertices can be written as $-kT^{a}g_{4}\gamma^{5}(d_{L,a}^{(m,n,p)}P_{L}+d_{R,a}^{(m,n,p)}P_{R})$ where
$k$ is the Lorentz breaking factor and
\be
d_{L/R,a}^{(m,n,p)}=\sqrt{2\pi R}\int_{0}^{2\pi R}\!\!\!\!\!dy\,f_{a,5}^{(p)}(y)\Big[f_{1,L/R}^{(n)}(y)f_{2,R/L}^{(m)}(y)+
\widetilde{f}_{1,L/R}^{(n)}(y)\widetilde{f}_{2,R/L}^{(m)}(y)\Big].
\label{c5LR}
\ee
The only coupling relevant for us is the one with the colored would-be Goldstone bosons $p=m=1$, $n=0$, for which one has
\be
\left| d_{L/R,g}^{(1,1,0)} \right|
=\frac{\rho _s}{k} \frac{m_{f_-}^{(1)}}{m_{g_-}^{(1)}}\,\,\left| c_{L/R,g}^{(1,1,0)} \right|\,\,\,\,.
\ee

\section{One--loop Gluon Mass Correction}

One-loop computations on orbifolds are conveniently performed by using the method of images
to map the propagators on $S^1/\Z_2$ to those on the covering circle $S^1$ \cite{Georgi:2000ks}.
In this way, the vertices conserve the KK number and the KK violation induced by the
boundaries is all encoded in a term in the propagator of the bulk fields.

As discussed in the main text, the only radiative correction of interest
to us is the mass splitting $\Delta m_{g}=m_{g^{(2)}}-2m_{g^{(1)}}$.
There are three classes of radiative corrections:
i) bulk (finite) corrections induced by bulk fields, ii) localized (divergent)
corrections induced by bulk fields and iii) localized (divergent) corrections
induced by boundary fermion fields. The corrections i) and ii) are one-to-one, in the formalism of \cite{Georgi:2000ks}, to loop corrections with respectively an even or odd number of insertions
of KK-violating propagator terms.

This picture is valid in the limit of vanishing bulk-to-boundary mixing terms
($\epsilon\rightarrow 0$), that is
a very good approximation for all the fermions but the top. In the latter
case, the calculations are more involved, since $\epsilon_t\sim O(1)$ and
the corrections ii) and iii) cannot be separated.
We have nevertheless checked that the effect of $\epsilon$ is negligible
in the radiative correction. Indeed, by taking the opposite limit $\epsilon\rightarrow\infty$,
in which several simplifications occur,
the top contribution to the mass splitting varies $\sim1\%$ with respect to the $\epsilon=0$ contribution.
For all practical purposes, we can thus safely take $\epsilon=0$ for all SM fields and consider separately
contributions ii) and iii).

\subsection{Bulk Contributions}

Bulk contributions are easily computed. Since there are no bulk fields charged under both
$SU(3)_{1,s}$ and $SU(3)_{2,s}$, mirror symmetry constrains the one-loop mass corrections to the gluons
$g_1$ and $g_2$ (and hence $g_+$ and $g_-$) to be equal.
Divergencies appear but they are associated with the renormalization of the 5D coupling
constant and the Lorentz violating parameter $\rho_s$. The former does not alter the mass spectrum
and the latter dependence clearly cancels in computing $\Delta m_{g}$. What is left is a finite
universal correction, similarly to the case of \cite{Cheng:2002iz}.
The purely bosonic and ghost contributions are as in \cite{Cheng:2002iz}, once one rescales
$1/R\rightarrow\rho_{s}/R$, since the Feynman rule for periodic and antiperiodic fields are
essentially the same. Antiperiodic odd fields running in the loop give only rise to
a phase $(-)^w$ after a Poisson resummation on the KK modes is peformed.
The contributions of virtual odd fields in the diagrams is proportional to
$\sum_{w=1}^\infty (-)^w/w^3=-3 \zeta(3)/4$, and equals then $(-3/4)$ times the ones of the
corresponding even fields, giving a partial cancellation. In total, the gluon and ghost contributions equal
\be
\delta m_{g^{(n)}}^2\Big|_{g.+gh.}=\frac{9}{8}\frac{\alpha_{s}\zeta(3)}{\pi^{3}}
\frac{\rho_{s}^{2}}{R^{2}}\Big(1-\frac{3}{4}\Big)\,.
\label{dmbulkbos}
\ee
Eq.~(\ref{dmbulkbos}) is valid for all periodic (even $n$) and antipeioridc (odd $n$) modes and is independent of the KK number of the external gluons, with the only exception of the $n=0$ massless QCD gluons for which one clearly
has $\delta m_{g^{(0)}}^2=0$ by gauge invariance.

Fermion loops are similarly treated, although now the Lorentz breaking cannot be simply
rescaled away. For a couple of fermion pairs $(\Psi_{1,2}\tilde\Psi_{1,2})$ in the fundamental
representation of $SU(3)_{1,2,s}$ with bulk mass $M_i$ and Lorentz breaking parameter $k_i$, one finds
\be
\delta m_{g^{(n)}}^{2}\Big|_{fer.}\simeq -\frac{\alpha_s k^2}{\pi^3 R^2} \sum_{w=1}^\infty
\frac{e^{-2 w \lambda_i/k_i}}{w^3}\Big(\frac{1+(-)^w}{2}\Big) \bigg[1+2w \frac{\lambda_i}{k_i}
\bigg]\,,
\label{dmbulkferm}
\ee
where $\lambda_i=\pi M_i R$ and we have neglected negligible corrections $O(1-k^2/\rho_s^2)$ in Eq.~(\ref{dmbulkferm}). The terms proportional to 1 and $(-)^w$ in Eq.~(\ref{dmbulkferm}) correspond
(for $\Z_2$ even gluons) to periodic and antiperiodic fermion contributions respectively. As above, a partial cancellation of the mass correction is induced by antiperiodic fields.
Again, the mass correction (\ref{dmbulkferm}) is valid for any KK number
of the external gluons, but the $n=0$ gluons.

\subsection{Localized Contributions from Bulk Fields}

Due to the presence of one non-diagonal propagator,
no sum over KK modes has to be performed in the Feynman diagram loop associated to these
contributions. The diagrams are effectively four dimensional and logarithmically divergent.
Such divergencies are cancelled by introducing boundary kinetic counterterms for the gluons
at the orbifold fixed points.
Strictly speaking, this kind of contributions would then be
uncalculable, depending on the arbitrary renormalization prescription chosen to
cancel these divergencies. It is however possible to estimate their effect by assuming
that they are dominated by the calculable radiative corrections of the model.
In other words, we require as renormalization prescription the vanishing
of these counterterms at a scale of energy equal to the cut-off $\Lambda$ of the theory.

The mass correction is encoded in the $\eta_{\mu\nu}$ coefficient $\Pi$
of the gluon vacuum polarization term, taken at $p^2=m_{g^{(n)}}^2$.
Contrary to the bulk terms, boundary corrections also induce mixing between the KK modes,
so that a diagonalization of an infinite mass matrix should be performed in order to
get the mass eigenvalues. All off--diagonal components are however one--loop induced,
so that at one--loop level we can safely neglect such terms and focus only on the diagonal
two--point amplitudes.
Since the $\Pi$ factor is given by a 4D loop diagram, its form is the same for
periodic and antiperiodic gluons. The only non-trivial issue is the sign of the mass
correction. The latter is fixed by the boundary conditions (\ref{bound-cond}).\footnote{Instead
of considering periodic and antiperiodic fields, as usual, one could alternatively consider an
$S^1/(\Z_2\times \Z_2^\prime)$ orbifold where all fields are periodic but with different orbifold
parities at $y=0$ and $y=\pi R$.} The ending result is that no localized mass term is induced
at $y=0$, whereas at $y=\pi R$ the periodic and antiperiodic contributions are equal.
The localized mass contributions induced by gluon and ghost fields is found to be ($n>0$)
\be
\delta m_{g^{(n)}}^{2}=\frac{23\alpha_{s}}{4\pi}m_{g^{(n)}}^2\ln\Big(\frac{\Lambda}{m_{g^{(n)}}}\Big)\,,
\label{boundary-loc}
\ee
where $m_{g^{(n)}}=\rho_{s} n/(2R)$ is the tree--level mass for periodic and antiperiodic gluons.
The localized contributions induced by bulk fermion fields vanish trivially because
the KK--violating terms in the fermion propagator contains a $\gamma^5$ factor which
results in a vanishing trace over the spinor indices. From Eq.~(\ref{boundary-loc}) we get the following
one-loop contribution to $\Delta m_g$:
\be
\Delta m_g = -\frac{23\alpha_{s}}{8\pi}\frac{\rho_s}{R}\ln 2\,,
\ee
independently of the cut--off $\Lambda$.

\subsection{Localized Contributions from Boundary Fields}

The contributions from colored fermions localized at $y=0$ is straightforward.
Being a purely 4D contribution, it is logarithmically divergent and will be
renormalized as described before, requiring the vanishing of the localized operator
at the scale $\Lambda$.\footnote{As we have seen, the operator induced by bulk fields
is localized only at $y=\pi R$ and thus the renormalization prescription performed
here is independent from the one of section B.2.}.
Boundary fermions do not minimally couple to $g^{(n)}_-$, so that $\delta m_{g^{(1)}}^{2}=0$.
Summing over all colored fields, for periodic KK gluons ($n> 0$) we find
\be
\delta m_{g^{(n)}}^{2}=-\frac{\alpha_{s}}{3\pi} m_{g^{(n)}}^{2}
\ln\Big(\frac{\Lambda}{m_{g^{(n)}}}\Big)\times 12 \,.
\ee

We summarize in Table~\ref{tabcorr} the different kind of contributions, summed
over all the fields in the model.

\begin{table}[htb]
\begin{center}
\begin{tabular}{|c|c|}
\hline
& $\Delta m_g$
\tabularnewline
\hline
\hline
\b
i) bulk bosons&
$-\frac{27\,\zeta(3)}{16\pi^{2}}$
\tabularnewline
\hline
\b
i) bulk fermions&
$\frac{3}{\pi^{2}}$\tabularnewline
\hline
\b
ii) bulk bosons&
$-\frac{23}{2}\ln(2)$
\tabularnewline
\hline
\b
ii) bulk fermions&
$0$\tabularnewline
\hline
\b
iii) boundary fermions&
$-8\ln\Big(\frac{\Lambda}{m_{g^{(2)}}}\Big)$\tabularnewline
\hline
\end{tabular}
\end{center}
\label{tabcorr}
\caption{Summary of mass corrections in terms of $\frac{\alpha_{s}}{4\pi}\frac{\rho_{s}}{R}$}
\end{table}

{}For a cut--off scale $\Lambda\simeq (3\div 4)/R$, the mass splitting
$\Delta m_g$ turns out to be approximately equal to
\be
\Delta m_{g}=m_{g_{+}}^{(2)}-2m_{g_{-}}^{(1)}\simeq   -1.4\, \alpha_{s}\frac{\rho_{s}}{R}.
\label{DMapp}
\ee

\begin{table}[h]
\small{\begin{tabular}{p{5.5cm}p{2cm}p{2cm}p{2.0cm}p{0.5cm}}
 & \multicolumn{4}{c}{Diagrams} \\ \cline{2-5}
Process & s  & t & u & 4p \\
\hline
~ & \\[-3.0ex]
$A^{(1)}_- A^{(1)}_- \into (f_R \bar f_R, f_{L} \bar f_{L})$ &  & \hspace{-.6cm} $(\chi_a,\psi_a)$ &
\hspace{-.6cm} $(\chi_a,\psi_a)$ \\
$A^{(1)}_- A^{(1)}_- \into  (b^{(0)}_+ \bar b^{(0)}_+,\tau^{(0)}_+ \bar \tau^{(0)}_+)$ & & $\phi_a$ & $\phi_a$ \\
\hline
~ & \\[-3.0ex]
$\chi_a  \bar \chi_a  \into q_R \bar q_R$ & $g^{(0)}_+ ,g^{(2)}_+ $& $g^{(1)}_- $& &   \\
$\chi_a  \bar \chi_a  \into q_L \bar q_L$ & $g^{(0)}_+ ,g^{(2)}_+ $& & &\\
$\chi_a  \bar \chi_a  \into b^{(0)}_+ \bar b^{(0)}_+$ & $g^{(0)}_+ $& & &\\
$\chi_a  \bar \chi_b  \into q_R \bar q_R$ & & $g^{(1)}_- $& &   \\
$\chi_a  \bar \chi_a  \into g^{(0)}_+ g^{(0)}_+$ & $g^{(0)}_+ $& $g^{(1)}_- $&$g^{(1)}_- $&  \\
$\chi_a  \chi_{a,b}  \into q_R q_R$ & & $g^{(1)}_- $&$g^{(1)}_- $&   \\
$\psi_a  \bar \psi_a  \into q_{L} \bar q_{L}$ & $g^{(0)}_+ ,g^{(2)}_+ $& $g^{(1)}_- $& &   \\
$\psi_a  \bar \psi_a \into q_R \bar q_R$ & $g^{(0)}_+ ,g^{(2)}_+ $& & &\\
$\psi_a  \bar \psi_a \into b^{(0)}_+ \bar b^{(0)}_+$ & $g^{(0)}_+ $& & &\\
$\psi_a  \bar \psi_b  \into q_L \bar q_L$ & & $g^{(1)}_- $& &   \\
$\psi_a  \bar \psi_a  \into g^{(0)}_+ g^{(0)}_+$ & $g^{(0)}_+ $& $g^{(1)}_- $&$g^{(1)}_- $&  \\
$\psi_a  \psi_{a,b}  \into q_{L} q_{L}$ & & $g^{(1)}_- $&$g^{(1)}_- $&   \\
$\phi_a  \bar \phi_a  \into b^{(0)}_+ \bar b^{(0)}_+$ & $g^{(0)}_+$& $g^{(1)}_- $&  \\
$\phi_a  \bar \phi_a  \into (q_R \bar q_R, q_{L} \bar q_{L})$ & $g^{(0)}_+,g^{(2)}_+ $& &  \\
$\phi_a  \bar \phi_b  \into b^{(0)}_+ \bar b^{(0)}_+$ & & $g^{(1)}_- $&  \\
$\phi_a  \bar \phi_a  \into g^{(0)}_+ g^{(0)}_+$ & $g^{(0)}_+ $& $g^{(1)}_- $&$g^{(1)}_- $&  \\
$\phi_a  \phi_{a,b}  \into b^{(0)}_+ b^{(0)}_+$ & & $g^{(1)}_- $&$g^{(1)}_- $&  \\
$\chi_{a,b}  \bar \psi^{\bnb}_{a,b}  \into q_R \bar q\bnbs_{L}$ & & $g^{(1)}_- $& &   \\
$\phi_{a,b}  \bar \psi^{\bnb}_{a,b}  \into b^{(0)}_+ \bar q\bnbs_L  $ & & $g^{(1)}_- $& &   \\
$\phi_{a,b}  \bar \chi\bnb_{a,b}  \into b^{(0)}_+ \,\bar q\bnbs_R $ & & $g^{(1)}_- $& &   \\
\hline
~ & \\[-3.0ex]
$A^{(1)}_- \chi_{a,b} \into g^{(0)}_+ q_R$ &$\chi_{a,b}$ & &$\chi_{a,b}$ \\
$A^{(1)}_- \psi_{a,b} \into g^{(0)}_+ q_{L} $ & $\psi_{a,b}$ & &$\psi_{a,b}$\\
$A^{(1)}_- \phi_{a,b} \into g^{(0)}_+ b^{(0)}_+ $ & $\phi_{a,b}$ & &$\phi_{a,b}$\\
\hline
~ & \\[-3.0ex]
$g^{(1)}_- g^{(1)}_- \into (q_R \bar q_R, q_{L} \bar q_{L})$ & $g^{(0)}_+ ,g^{(2)}_+ $& \hspace{-.5cm} $(\chi_a,\psi_a)$ & \hspace{-.5cm} $(\chi_a,\psi_a)$ \\
$g^{(1)}_- g^{(1)}_- \into b^{(0)}_+ \bar b^{(0)}_+$ & $g^{(0)}_+ $& $\phi_a$ & $\phi_a$ \\
$g^{(1)}_- g^{(1)}_- \into g^{(0)}_+ g^{(0)}_+ $ &$g^{(0)}_+  $ & $g^{(1)}_- $ & $g^{(1)}_-$ & x \\
\hline
~ & \\[-3.0ex]
$A^{(1)}_- g^{(1)}_- \into (q_R  \bar q_R, q_{L} \bar q_{L})$ & & \hspace{-.5cm}$(\chi_a,\psi_a)$ & \hspace{-.5cm} $(\chi_a,\psi_a)$ \\
$A^{(1)}_- g^{(1)}_- \into b^{(0)}_+ \bar b^{(0)}_+$ & &$\phi_a$ & $\phi_a$ \\
\hline
~ & \\[-3.0ex]
$g^{(1)}_- \chi_{a,b} \into g^{(0)}_+ q_R $ &$\chi_{a,b}$ &$g^{(1)}_- $&$\chi_{a,b}$ \\
$g^{(1)}_- \psi_{a,b} \into g^{(0)}_+ q_{L} $ & $\psi_{a,b}$ &$g^{(1)}_- $ &$\psi_{a,b}$\\
$g^{(1)}_- \phi_{a,b} \into g^{(0)}_+ b^{(0)}_+ $ & $\phi_{a,b}$ &$g^{(1)}_- $ &$\phi_{a,b}$\\
\hline
~ & \\[-3.0ex]
\end{tabular}}
\caption{List of all the relevant (co--)annihilation processes. See text for details.
}
\label{tab:coanns}
\end{table}

\section{Annihilation and coannihilation processes}

We collect in Table~\ref{tab:coanns} all the matrix elements which are relevant
for the computation of the DM relic density. Recall that the bulk fermions are in either
the ${\bf \bar 3}_{1/3}$ or ${\bf 6}_{1/3}$ of $SU(3)_w$, where in the subscript we have
denoted their $U(1)$ charge under $U(1)_+$. After EWSB, they decompose as follows
under $SU(2)_L\times U(1)_Y$: ${\bf \bar 3}_{1/3} = {\bf 2}_{1/6} \oplus {\bf 1}_{2/3}$ and
${\bf 6}_{1/3} = {\bf 3}_{2/3}\oplus {\bf 2}_{1/6} \oplus {\bf 1}_{-1/3}$.
In Table~\ref{tab:coanns} we have denoted by  $\chi$, $\psi$ and $\phi$ respectively the
$SU(2)_L$ singlet, doublet and triplet components of the lightest $n=1$ KK mode of the 5D antiperiodic
bulk fermions $\Psi_-$ in both the ${\bf \bar 3}$ and the ${\bf 6}$, with the
understanding that for the ${\bf \bar 3}$ $\phi$ (and the corresponding processes) are missing.
These fields coincide with the states that we have collectively denoted by $b^{(1)}_-$, $c^{(1)}_-$, etc.
in Fig.~\ref{figmass} and in the main text.
The subscript $a,b=1,2$ refers to the two distinct towers of KK mass eigenstates coming from the
fermion pairs $(\Psi_-,\tilde \Psi_-)$. The SM fermions are denoted by $f$ when we are considering
both quarks and leptons and $q$ for quarks only. We denoted by $b^{(0)}_+$ and $\tau_+^{(0)}$
the $n=0$ KK mode of the $SU(2)_L$ periodic triplets arising from the 5D bulk fermions
$\Psi_+^{b,\tau}$, as in Fig.~\ref{figmass}.
For each process, we also write the particle exchanged in the various $(s,t,u)$ channels,
whenever the flavour and gauge symmetries allow it. The channels mediated by $g^{(1)}_-$ should be
considered only for the framework of Section 3.2.
The fourth column $4p$ indicates when a four-point interaction vertex is present.

\begin{table}[!h]
\begin{center}
\begin{tabular}{|c|c|c|c|c|c|}
\hline
State& $A_-^{(1)}$&$g_-^{(1)}$&$b_-^{(1)}$&$c_-^{(1)}$&$\tau_-^{(1)}$
\tabularnewline
\hline
\hline
\b
$D.F.$&$3$&$24$&$144$&$72$&$48$
\tabularnewline
\hline
\end{tabular}
\end{center}
\caption{Degrees of freedom for the states involved in coannihilation.}
\label{tabdf}
\end{table}

In Table~\ref{tabdf} we list the degrees of freedom for the states relevant in the computation of the $A_-^{(1)}$ relic abundance.
For fermions we have $D.F.=2\times 4\, N_c n_s$, where $N_c$ is the color factor and $n_s$ the number of states in the $SU(3)_w$ multiplet.
The overall factor $2$ takes into account the presence of two distinct towers for the antiperiodic fermions.
In the case of gauge bosons one has simply $D.F.= 3 N_g$, where $N_g$ is the number of generators of the gauge group.

\section{Running of $\alpha_s$}

The effective annihilation rate of our DM candidate is dominated
by coannihilation and resonance effects with colored particles. As a result, the DM abundance
is quite sensitive to the strong coupling constant $\alpha_s$: $\Omega_{DM}\propto \alpha_s^{-2}$, where
$\alpha_s$ is here evaluated at the typical energy scale of the annihilation processes, namely $1/R$.
It is then important to evolve $\alpha_s$ from, say, the scale of the $Z$ boson mass $m_Z$
up to $1/R$ taking into account the various threshold effects due to the many
particle states in the model with a mass in that range. Due to the limited energy range of the
RG flow and the presence of several states with a mass close to $1/R$, we have computed the
QCD $\beta$--function at one--loop level in a mass dependent scheme (momentum subtraction) \cite{Georgi:1976ve},
rather than in the more usual
Minimal Subtraction (MS) or modified MS ($\overline{\rm MS}$) schemes. In this way, all the effects
of threshold corrections are automatically taken into account, with no need
of matching conditions. The drawback, of course, is that the
resulting $\beta$--function has a much more complicated form and the RG evolution has to be performed numerically.

The computation is considerably simplified in a background field gauge (with $\xi=1$ gauge fixing parameter), where the $\beta$--function is extracted by considering only the gluon 2--point function.
The contribution of a Dirac fermion and a gauge field with mass $M$ in a representation $r$ of $SU(3)_s$ can
be written in a compact way as $\beta(g_s,M/\mu) = g_s(\mu)^3/16\pi^2 [\beta_f(M/\mu)+\beta_g(M/\mu)]$, with\footnote{Notice
that for unperturbed KK states (i.e. with vanishing bulk--to--boundary mixing terms), the $\beta$--function contribution
of the whole tower of KK states can be resummed, giving rise to $\mu R$--dependent Coth or Tanh hyperbolic functions (for periodic or antiperiodic fields, respectively). One can then easily check that for $R\rightarrow 0$ or $R\rightarrow \infty$ these functions interpolate between the usual 4D log--regime and the 5D linear--regime for $\alpha_s$. One should however recall that in our model $\Lambda \simeq (3\div 4)/R$ so that for $\mu\sim \Lambda$ we enter in a strong coupling regime where our perturbative
expressions break down.}
\bea
\beta_f(M/\mu)& = & 8 T_f(r) \int_0^1 \! \frac{\mu^2x^2(1-x)^2}{M^2+\mu^2x(1-x)}\,,\label{betaFer} \\
\beta_g(M/\mu) & = & -T_g(r) \int_0^1 \! \frac{\mu^2x(1-x)(1+9x-6x^2)}{M^2+\mu^2x(1-x)} \label{betaGauge} \,,
\eea
and the convention that $T(fund.)=1/2$. Eqs.(\ref{betaFer}) and (\ref{betaGauge}) give the correct contributions
for $M\rightarrow 0$: $\beta_f(g)\rightarrow 4/3 T_f(r)$ and $\beta_g(g)\rightarrow T_g(r) (-7/2= -11/3+1/6)$,
where $-11/3$ and $1/6$ are the contributions of the transverse and logitudinal (scalar) components of the gauge field.
Eq.~(\ref{betaFer}) is also in agreement with \cite{Georgi:1976ve}, whereas we are not aware of the presence
of an explicit expression for $\beta_g$ in the literature. At one--loop level, one can consistently take the masses
to be constant and $\mu$--independent. Given the above relation, we have then numerically computed the RG evolution
of $\alpha_s$ starting from its value $\alpha_s(M_Z)=0.117$ up to $1/R$, by including all particles in the model (for all states $T_f=T_f(fund.)=1/2$ and $T_g=T_g(adj.)=3$) with a mass up to $4/R$, to take also into account of residual
threshold effects of more massive states.
For $\mu\lesssim 1/(2R)$, $\alpha_s$ decreases but then, in the range $1/(2R)\leq \mu\leq 1/R$,
due to the many colored fermions which become active, the sign of $\beta$ changes and $\alpha_s$ starts to increase.
The two effects tend to compensate each other, so that eventually $\alpha_s(1/R)$ is quite close
to $\alpha_s(M_Z)$. For example, for $1/R\simeq 5$ TeV, $\alpha_s(1/R)\simeq 0.097$ and $\alpha_s(1/R)\simeq 0.092$ for the model with one or two copies of $SU(3)_s$, respectively.
Notice, just for comparison, that in a UED model with integer KK modes and $1/R\simeq 2.5$ TeV, we would have got
a value significantly lower for $\alpha_s$: $\alpha_s(2/R)\simeq 0.075$.

\end{document}